\documentclass[fleqn,usenatbib,useAMS]{mnras}

\usepackage{newtxtext,newtxmath}

\usepackage[T1]{fontenc}
\usepackage{ae,aecompl}

\DeclareRobustCommand{\VAN}[3]{#2}
\let\VANthebibliography\thebibliography
\def\thebibliography{\DeclareRobustCommand{\VAN}[3]{##3}\VANthebibliography}

\usepackage{graphicx}	
\usepackage{amsmath}	

\usepackage{amssymb}	
\usepackage{array}
\usepackage{multirow}
\usepackage{tabularray}
\usepackage{subcaption} 
\usepackage{multirow}
\usepackage{hyperref}
\usepackage{multicol}        
\usepackage{bm}		
\usepackage{pdflscape}	
\usepackage{siunitx}
\usepackage{threeparttable}
\usepackage{lscape}
\usepackage{rotating}
\usepackage{xcolor}

\newcommand{\angstrom}{\textup{\AA}}

\title[Quasar catalogue and composite spectrum]{Optical and near-infrared spectroscopy of quasars at $z>6.5$: public data release and composite spectrum}

\author[Onorato et al.]{
Silvia Onorato$^{1}$\thanks{E-mail: onorato@strw.leidenuniv.nl},
Joseph F. Hennawi$^{1,2}$,
Jan-Torge Schindler$^{3}$,
Jinyi Yang$^{4}$,
Feige Wang$^{4}$, \newauthor
Aaron J. Barth$^{5}$,
Eduardo Ba{\~n}ados$^{6}$,
Anna-Christina Eilers$^{7}$,
Sarah E. I. Bosman$^{6,8}$, \newauthor
Frederick B. Davies$^{6}$, 
Bram P. Venemans$^{1}$,
Chiara Mazzucchelli$^{9}$,
Silvia Belladitta$^{6,10}$, \newauthor
Fabio Vito$^{10}$, 
Emanuele Paolo Farina$^{11}$,
Irham T. Andika$^{12,13}$,
Xiaohui Fan$^{14}$, \newauthor
Fabian Walter$^{6}$, 
Roberto Decarli$^{10}$,
Masafusa Onoue$^{15,16,17}$,
and Riccardo Nanni$^{1}$
\\
$^{1}$Leiden Observatory, Leiden University, P.O. Box 9513, NL-2300 RA Leiden, the Netherlands\\
$^{2}$Department of Physics, Broida Hall, University of California, Santa Barbara, Santa Barbara, CA 93106-9530, USA\\
$^{3}$Hamburger Sternwarte, Universität Hamburg, Gojenbergsweg 112, D-21029 Hamburg, Germany\\
$^{4}$Department of Astronomy, University of Michigan, 1085 S. University Ave., Ann Arbor, MI 48109, USA\\
$^{5}$Department of Physics and Astronomy, 4129 Frederick Reines Hall, University of California, Irvine, CA 92697-4575, USA\\
$^{6}$Max Planck Institut f\"ur Astronomie, K\"onigstuhl 17, D-69117, Heidelberg, Germany\\
$^{7}$MIT Kavli Institute for Astrophysics and Space Research, 77 Massachusetts Avenue, Cambridge, 02139, Massachusetts, USA\\
$^{8}$Institute for Theoretical Physics, Heidelberg University, Philosophenweg 12, D-69120, Heidelberg, Germany\\
$^{9}$Instituto de Estudios Astrof\'{\i}sicos, Facultad de Ingenier\'{\i}a y Ciencias, Universidad Diego Portales, Avenida Ejercito Libertador 441, Santiago, Chile\\
$^{10}$INAF -- Osservatorio di Astrofisica e Scienza dello Spazio di Bologna, Via Gobetti 93/3, I-40129 Bologna, Italy\\
$^{11}$Gemini Observatory, NSF’s NOIRLab, 670 N A’ohoku Place, Hilo, Hawai'i 96720, USA\\
$^{12}$Technical University of Munich, TUM School of Natural Sciences, Department of Physics, James-Franck-Str. 1, D-85748 Garching, Germany\\
$^{13}$Max-Planck-Institut f\"{u}r Astrophysik, Karl-Schwarzschild-Str. 1, D-85748 Garching, Germany\\
$^{14}$Steward Observatory, University of Arizona, 933 N. Cherry Ave, Tucson, AZ 85721, USA\\
$^{15}$Kavli Institute for the Physics and Mathematics of the Universe (Kavli IPMU, WPI), The University of Tokyo Institutes for Advanced Study,\\
The University of Tokyo, Kashiwa, Chiba 277-8583, Japan\\
$^{16}$Center for Data-Driven Discovery, Kavli IPMU (WPI), UTIAS, The University of Tokyo, Kashiwa, Chiba 277-8583, Japan\\
$^{17}$Kavli Institute for Astronomy and Astrophysics, Peking University, Beijing 100871, P.R.China}

\date{Accepted 2025 May 12. Received 2025 May 9; in original form 2024 June 11}

\pubyear{2025}

\begin{document}
\label{firstpage}
\pagerange{\pageref{firstpage}--\pageref{lastpage}}
\maketitle

\begin{abstract}
We present optical and near-infrared (NIR) spectroscopic observations for a sample of $45$ quasars at $6.50 < z \leq 7.64$ with absolute magnitudes at $1450$ {\angstrom} in the range $-28.82 \leq M_{1450} \leq -24.13$ and their composite spectrum. The median redshift and $M_{1450}$ of the quasars in the sample are $z_{\rm{median}}=6.71$ and $M_{1450,\rm{median}} \simeq -26.1$, respectively.
The NIR spectra are taken with echelle spectrographs, complemented with additional data from optical long slit instruments, and then reduced consistently using the open-source Python-based spectroscopic data reduction pipeline \texttt{PypeIt}.
The median of the mean signal-to-noise ratios per $110$ km s$^{-1}$ pixel in the J, H, and K band [median $\langle \rm{SNR}_{\lambda} \rangle$] is: median $\langle \rm{SNR}_{J} \rangle=9.7$, median $\langle \rm{SNR}_{H} \rangle=10.3$, and median $\langle \rm{SNR}_{K} \rangle=11.7$; demonstrating the good data quality.
This work presents the largest medium/moderate-resolution sample of quasars at $z>6.5$ from ground-based instruments. 
Despite the diversity in instrumental setups and spectral quality, the data set is uniformly processed and well-characterized, making it ideally suited for several scientific goals, including the study of the quasar proximity zones and damping wings, the Ly$\alpha$ forest, the intergalactic medium's metal content, as well as other properties such as the distribution of SMBH masses and Eddington ratios. Our composite spectrum is compared to others at both high and low-$z$ from the literature, showing differences in the strengths of many emission lines, probably due to differences in luminosity among the samples, but a consistent continuum slope, which proves that the same spectral features are preserved in quasars at different redshift ranges.
\end{abstract}




\section{Introduction}\label{sec:intro}
Discoveries of high-redshift quasars ($z > 6$) have uncovered the presence of exceptionally supermassive black holes (SMBHs), ranging from approximately $10^8$ to $10^{10}$ $M_{\odot}$, in the early stages of the Universe (\citealt{Wu2015}; \citealt{Banados2018}; \citealt{Matsuoka2019b}; \citealt{Onoue2019}; \citealt{Shen2019}; \citealt{Yang2020a}; \citealt{Wang2021a}; see \citealt{Fan2023} for a recent review). This motivates inquiry into the rapid growth mechanisms enabling these SMBHs to reach billions of solar masses within an extraordinarily brief time frame, less than one billion years after the big bang.

Various theoretical models proposing different seed black hole masses ($10^2-10^{6}$ $M_{\odot}$) and highly efficient accretion modes present potential explanations for the formation and growth of early SMBHs (see \citealt{Woods2019} and \citealt{Inayoshi2020} for recent reviews). 
Rigorous observations of an extensive sample of $z>6.5$ quasars are imperative to test and refine these models, advancing our comprehension of SMBH formation and evolution.
Recent advancements in deep imaging surveys, coupled with the enhanced capabilities of near-infrared (NIR) spectroscopy on large telescopes, have substantially expanded the sample size of quasars with $z > 6$ to more than $200$ (e.g., \citealt{Fan2023}). 
This has extended the quasar frontier beyond $z > 7.5$, moving deep into the epoch of reionization (EoR; e.g., \citealt{Mortlock2011}; \citealt{Jiang2016}; \citealt{Mazzucchelli2017}; \citealt{Banados2016, Banados2018}; \citealt{Fan2019}; \citealt{Reed2019}; \citealt{Matsuoka2019b, Matsuoka2019a}; \citealt{Venemans2013, Venemans2015}; \citealt{Wang2018, Wang2019}; \citealt{Yang2019, Yang2020a, Yang2021}; \citealt{Wang2021a}).
Furthermore, recent observations from the \textit{James Webb Space Telescope} (JWST) have already uncovered new high-$z$ active galactic nuclei (AGNs) with lower luminosity and black hole mass, promising to deliver additional breakthroughs in our understanding (e.g., \citealt{Labbe2023}; \citealt{Maiolino2023a, Maiolino2023b}; \citealt{Larson2023}; \citealt{Fujimoto2023}; \citealt{Goulding2023}; \citealt{Furtak2023}; \citealt{Kokorev2023}; \citealt{Greene2023}; \citealt{Kokorev2024}; \citealt{Perez-Gonzalez2024}; \citealt{Matthee2024}; \citealt{Andika2024}). 

Our work of collecting high-quality spectra marks a significant milestone as the most extensive medium/moderate-resolution collection of quasar data at very high redshifts obtained through ground-based instruments.
Despite the diversity in spectral resolution and signal-to-noise ratios (SNR), the data set is consistently processed, ensuring reproducibility and reliability for various scientific objectives. These include the exploration of the final phases of the EoR through the study of the Ly$\alpha$ forest \citep{Fan2006, Becker2015, Bosman2022}, investigations into the intergalactic medium's (IGM) metal content (e.g., \citealt{Davies2023}), the broad-line region chemical abundances (e.g., \citealt{Lai2022}), and analyses of properties such as the distribution of SMBH masses and Eddington ratios \citep{Yang2021, Farina2022, Mazzucchelli2023}.
Also, the creation of a composite spectrum can be extremely helpful in evaluating the continuum and line properties of quasar spectra, investigating the average rest-frame UV quasar spectral properties and their possible evolution with redshift (\citealt{VandenBerk2001}; \citealt{Selsing2016}; \citealt{Mazzucchelli2017}; \citealt{Meyer2019}; \citealt{Schindler2020}; \citealt{Yang2021}). 
Moreover, constructing a composite spectrum enhances the detection of spectral features that may be difficult to identify in individual spectra due to their low SNR. This approach also facilitates the identification of objects that deviate from the template and aids in modelling and fitting spectral energy distributions.

Other key analyses involving high-quality spectra aim to place constraints on the average fraction of neutral hydrogen ($\langle x_{\rm{HI}} \rangle$) at the EoR and on the radiative efficiency of the earliest SMBHs. They can be performed by starting with the reconstruction of the quasar's intrinsic blue side of the spectrum from the observed red side employing, e.g., principal component analysis (PCA) continuum modelling (e.g., \citealt{Davies2018a, Davies2018b, Davies2019}; \citealt{Bosman2021}) so that we can systematically study quasar proximity zones and lifetimes $t_{\rm{Q}}$ (see \citealt{Fan2006}; \citealt{Eilers2017, Eilers2020}; \citealt{Davies2020}; \citealt{Satyavolu2023}), proximate damped Ly$\alpha$ systems (see \citealt{Banados2019}; \citealt{Andika2020,Andika2022}), and Ly$\alpha$ damping wings (\citealt{Miralda-Escude1998}; \citealt{Davies2018a}; \citealt{Durovcikova2024}; \citealt{Greig2024}).

This paper is the first of a series aimed at exploiting the wealth of information that could be obtained from this sample (e.g., the study of the quasar proximity zone sizes in Paper II, \citealt{Onorato2025}). Here, we present and make publicly available the optical and NIR spectra of $45$ quasars at $6.50 < z \leq 7.64$, and the composite rest-frame UV/optical spectrum.
We describe the data set including the quasar sample, the instruments, and their properties, with the data reduction in Section \ref{sec:dataset}.
Section \ref{sec:qsoprop} details the spectral calibration procedure and presents the main properties of the sample. We show a comparison between this sample and other spectroscopic data releases of high-$z$ quasars in Section \ref{sec:qsosamples}.
We discuss the mean quasar composite spectrum in Section \ref{sec:composite}, and we conclude this work with a summary of the paper, presented in Section \ref{sec:summary}.
All results below refer to a $\Lambda$CDM cosmology with parameters $\Omega_{\Lambda} = 0.7$, $\Omega_{m} = 0.3$, and $h = 0.7$; all magnitudes are reported in the AB system.

\section{The Data Set}\label{sec:dataset}
\subsection{Quasar Sample}\label{sec:sample}
We start this analysis by compiling a list of quasars, both known from the literature and unpublished, and applying a redshift cut at $z>6.5$. 
We check the individual telescope archives for medium/moderate-resolution spectroscopy of these sources from echelle spectrographs, and then also long slit instruments if the Ly$\alpha$ region is not sufficiently covered. 
We avoid collecting data of faint objects ($M_{1450} \gtrsim -25$), even if we do not apply a strict $M_{1450}$ cut as long as the luminosity of the source does not strongly compromise the data reduction.
Thus, our catalog includes $45$ quasars in the redshift range $6.50 < z \leq 7.64$ ($z_{\rm{median}}=6.71$) and in the magnitude range $-28.82 \leq M_{1450} \leq -24.13$ ($M_{1450,\rm{median}} \simeq -26.1$, see Figure \ref{fig:M1450} and Table \ref{tab:sample}). The full names of the quasars are in Table \ref{tab:sample}, while short names are adopted throughout the paper. Out of the $45$ objects, 11 are classified in the literature as Broad Absorption Lines (BAL) quasars, defined as those with absorption lines with FWHM $\gtrsim$ 2000 km s$^{-1}$. These BALs sources are: J0313$-$1806, J0038$-$1527, J1243$+$0100\footnote{Identified as a possible BAL in the discovery paper \citep{Matsuoka2019b}, but the spectroscopy quality does not lead to a clear classification.}, J0839$+$3900, J2348$-$3054, J0246$-$5219, J0910$-$0414, J0923$+$0402, J0706$+$2921, J1526$-$2050, and J0439$+$1634, as flagged in Table \ref{tab:sample}. We include three new unpublished high-$z$ quasars, J0410$-$0139 \citep{Banados2025}, J1917$+$5003, and J0430$-$1445 \citep{Belladitta2025}. This last source also seems to be a possible BAL, leading to a final number of $12$ BALs in this work and hence a BAL fraction of $\simeq 27\%$. The fraction of BAL quasars measured by \cite{Bischetti2022} in the XQR-30 sample ($5.8 \lesssim z \lesssim 6.6$) is $\simeq 40\%$. They point out that other works collecting $z \gtrsim 5.7$ quasar spectra found a BAL fraction of $16-24\%$ \citep{Shen2019, Schindler2020, Yang2021}, and suggest these could be lower limits on the actual BAL fraction, as they are based on spectra with lower resolution and SNR. Similarly, our BAL fraction may also be a lower limit due to the heterogeneous data quality. Table \ref{tab:excluded} lists the remaining \textit{published} $z>6.50$ sources not included in the sample, stating for each quasar their redshift and the reason for exclusion (i.e., $M_{1450} \gtrsim -25$ or no echelle data).

\begin{landscape}
    \begin{table}
        \centering
        \begin{threeparttable}
	\caption{Information on the $45$ quasars in this sample, sorted by decreasing $z$. The details on the columns are provided at the end of Section \ref{sec:snr}.}
	\label{tab:sample}
	\begin{tabular}{lcccccccccccc}
		\hline
		Name & Instrument (arms) & $\rm{t_{exp} (s)}$ & $z$ & $z_{\rm{method}}$ & $z_{\rm{Ref}}$ & $\rm{J_{AB}}$ & $M_{1450}$ & Discovery & $\langle \rm{SNR_{J}} \rangle$ & $\langle \rm{SNR_{H}} \rangle$ & $\langle \rm{SNR_{K}} \rangle$ \\
		\hline
            J031343.839$-$180636.404$^{\rm{a}}$ & GNIRS/NIRES & 27300/15840 & 7.6423 & [\ion{C}{II}] & 1 & 20.92 $\pm$ 0.13 & $-$26.208 & 1 & 15.9 & 18.8 & 27.2 \\
            J134208.110$+$092838.610 & X-Shooter (NIR) & (81600) & 7.5413 & [\ion{C}{II}] & 2 & 20.64 $\pm$ 0.08 & $-$26.336 & 3 & 33.5 & 43.7 & 13 \\
            J100758.264$+$211529.207 & GNIRS/NIRES & 21000/7920 & 7.5149 & [\ion{C}{II}] & 4 & 20.20 $\pm$ 0.18 & $-$26.818 & 4 & 18.5 & 19.5 & 24.2 \\
            J112001.480$+$064124.300 & X-Shooter (NIR/VIS) & (114000/113440) & 7.0851 & [\ion{C}{II}] & 2 & 20.17 $\pm$ 0.07 & $-$26.565 & 5 & 44.7 & 53.7 & 308.7 \\
            J124353.930$+$010038.500$^{\rm{a}}$ & GNIRS & 10944 & 7.07 & \ion{Mg}{II} & 26 & 23.57 $\pm$ 0.08$^{\rm{b}}$ & $-$24.130$^{\rm{d}}$ & 26 & 1.4 & 1 & 1.3 \\
            J003836.097$-$152723.636$^{\rm{a}}$ & GNIRS/X-Shooter (NIR/VIS) & 15300/(15600/17912) & 7.0340 & [\ion{C}{II}] & 6 & 19.69 $\pm$ 0.07 & $-$27.030 & 7 & 24.8 & 24.6 & 22.6 \\
            J025216.640$-$050331.810 & X-Shooter (NIR/VIS) & (28800/31200) & 7.0006 & [\ion{C}{II}] & 6 & 20.19 $\pm$ 0.07 & $-$26.625 & 8 & 29.1 & 39.7 & 12.7 \\
            J041009$-$013919$^{\rm{e}}$ & NIRES & 9360 & 6.9964 & [\ion{C}{II}] & 27 & 20.75 $\pm$ 0.07 & $-$25.858 & 27 & 6 & 5.9 & 8.1 \\
            J083946.880$+$390011.440$^{\rm{a}}$ & GNIRS & 16800 & 6.9046 & \ion{Mg}{II} & 9 & 20.39 $\pm$ 0.20 & $-$26.214 & 10 & 27.2 & 29.9 & 43.9 \\
            J234833.340$-$305410.000$^{\rm{a}}$ & X-Shooter (NIR/VIS) & (9200/8783) & 6.9018 & [\ion{C}{II}] & 11 & 21.10 $\pm$ 0.08 & $-$25.224 & 12 & 7.5 & 10.3 & 4.2 \\
            J024655.902$-$521949.950$^{\rm{a}}$ & X-Shooter (NIR/VIS) & (24000/24000) & 6.8876 & [\ion{C}{II}] & 6 & 21.20 $\pm$ 0.14 & $-$25.301 & 8 & 6.1 & 11.7 & 4.9 \\
            J002031.470$-$365341.800 & X-Shooter (NIR/VIS) & (4800/4800) & 6.861 & [\ion{C}{II}] & 29 & 20.42 $\pm$ 0.10 & $-$25.999 & 13 & 6 & 8.3 & 3.6 \\
            J191729.984$+$500313.540$^{\rm{e}}$ & NIRES/MODS & 12720/26400 & 6.853 & \ion{Mg}{II} & 28 & 20.66 $\pm$ 0.05 & $-$26.208 & 28 & 5.2 & 5.4 & 5.3 \\
            J221100.601$-$632055.845 & X-Shooter (NIR/VIS) & (36000/37961) & 6.8449 & [\ion{C}{II}] & 6 & 21.27 $\pm$ 0.18 & $-$25.470 & 8 & 9 & 12.9 & 4.8 \\
            J031941.660$-$100846.000 & NIRES/GMOS & 18750/15300 & 6.8275 & [\ion{C}{II}] & 6 & 20.88 $\pm$ 0.30 & $-$25.480 & 8 & 4.9 & 6.1 & 12.2 \\
            J041128.628$-$090749.700 & NIRES/MODS & 5760/30000 & 6.8260 & [\ion{C}{II}] & 6 & 20.02 $\pm$ 0.14 & $-$26.490 & 10 & 17.4 & 18 & 19.6 \\
            J112925.368$+$184624.330 & X-Shooter (NIR/VIS) & (12000/12000) & 6.823 & \ion{Mg}{II} & 9 & 20.90 $\pm$ 0.11 & $-$25.421 & 14 & 6.6 & 8.2 & 3.2 \\
            J010953.130$-$304726.300 & X-Shooter (NIR/VIS) & (21600/21600) & 6.7909 & [\ion{C}{II}] & 11 & 21.28 $\pm$ 0.14 & $-$25.200 & 12 & 6.7 & 9.3 & 3.7 \\
            J082931.979$+$411740.870 & GNIRS & 12600 & 6.773 & \ion{Mg}{II} & 9 & 20.26 $\pm$ 0.15 & $-$26.154 & 10 & 12.2 & 9.2 & 13.2 \\
            J021847.040$+$000715.200 & NIRES/LRIS & 5760/3583 & 6.7700 & [\ion{C}{II}] & 6 & 21.08	$\pm$ 0.30 & $-$25.896 & 9,15 & 2.8 & 3.3 & 3.6 \\
            J110421.580$+$213428.850 & GNIRS & 7200 & 6.7662 & [\ion{C}{II}] & 6 & 19.91 $\pm$ 0.11 & $-$26.506 & 10 & 16.1 & 14.3 & 17.6 \\
            J091013.651$+$165630.180 & GNIRS & 13200 & 6.7289 & [\ion{C}{II}] & 6 & 21.06 $\pm$ 0.13 & $-$25.346 & 10 & 8.9 & 7.3 & 10.7 \\
            J043043.660$-$144541.210$^{\rm{a,e}}$ & GNIRS & 12000 & 6.7142 & \ion{Mg}{II} & 28 & 20.78 $\pm$ 0.18 & $-$25.656 & 28 & 8.4 & 7.5 & 11.7 \\
            J083737.830$+$492900.600 & GNIRS & 15600 & 6.702 & \ion{Mg}{II} & 9 & 20.21 $\pm$ 0.17 & $-$26.069 & 10 & 26.4 & 23.7 & 36.2 \\
            J200241.594$-$301321.690 & GNIRS & 3600 & 6.6876 & [\ion{C}{II}] & 6 & 19.97 $\pm$ 0.16 & $-$26.622 & 9 & 9.7 & 6.4 & 8 \\
            J092358.997$+$075349.107 & GNIRS & 7200 & 6.6817 & [\ion{C}{II}] & 6 & 21.25 $\pm$ 0.26$^{\rm{b}}$ & $-$25.687 & 9 & 3.2 & 2 & 2.9 \\
            J104819.090$-$010940.210 & X-Shooter (NIR/VIS) & (4800/4800) & 6.6759 & [\ion{C}{II}] & 16 & 20.87 $\pm$ 0.06 & $-$25.864 & 17 & 5.1 & 6.7 & 2.9 \\
            J223255.150$+$293032.040 & GNIRS/X-Shooter (NIR/VIS) & 4800/(14400/14400) & 6.666 & [\ion{C}{II}] & 31 & 20.46 $\pm$ 0.12 & $-$25.831 & 18 & 10.6 & 9.2 & 8.5 \\
            J210219.230$-$145853.860 & GNIRS/NIRES & 10200/5760 & 6.6645 & [\ion{C}{II}] & 6 & 21.15 $\pm$ 0.20 & $-$25.421 & 10 & 9.4 & 8.5 & 10.4 \\
            J121627.584$+$451910.675 & GNIRS & 3600 & 6.66 & \ion{Mg}{II} & 9 & 21.02 $\pm$ 0.13 & $-$25.518 & 10 & 4.8 & 3.6 & 5.3 \\
            J091054.535$-$041406.846$^{\rm{a}}$ & GNIRS/NIRES & 3600/3600 & 6.6363 & [\ion{C}{II}] & 6 & 20.23 $\pm$ 0.07 & $-$26.397$^{\rm{f}}$ & 10 & 9.8 & 11.4 & 15.6 \\
            J092347.117$+$040254.580$^{\rm{a}}$ & X-Shooter (NIR/VIS) & (43200/43200) & 6.6330 & [\ion{C}{II}] & 6 & 20.14 $\pm$ 0.08 &  $-$26.524$^{\rm{f}}$ & 10,19 & 24.4 & 48.5 & 23.1 \\
            J002429.772$+$391318.980 & GNIRS & 13800 & 6.621 & [\ion{C}{II}] & 20 & 21.28 $\pm$ 0.48 & $-$25.316 & 21 & 8.5 & 6.1 & 9.7 \\
            J030516.920$-$315056.000 & X-Shooter (NIR/VIS) & (16800/16800) & 6.6139 & [\ion{C}{II}] & 30 & 20.60 $\pm$ 0.05 & $-$25.690 & 12 & 12.3 & 15.6 & 6.3 \\
            J070626.379$+$292105.460$^{\rm{a}}$ & NIRES/DEIMOS & 14960/17800 & 6.6037 & [\ion{C}{II}] & 6 & 19.14 $\pm$ 0.05 & $-$27.410 & 10 & 24.3 & 25.6 & 34.2 \\
            J213233.190$+$121755.260 & X-Shooter (NIR/VIS) & (33600/35809) & 6.5881 & [\ion{C}{II}] & 20 & 19.74 $\pm$ 0.03 & $-$26.914 & 20 & 30.6 & 35.3 & 16.5 \\
            J152637.840$-$205000.660$^{\rm{a}}$ & X-Shooter (NIR/VIS) & (43200/46234) & 6.5864 & [\ion{C}{II}] & 16 & 19.60 $\pm$ 0.08 & $-$27.030 & 20 & 33 & 45.6 & 18.9 \\
            J233807.032$+$214358.170 & GNIRS & 4500 & 6.586 & [\ion{C}{II}] & 29 & 20.75 $\pm$ 0.30 & $-$25.816 & 9 & 3.4 & 2.4 & 2.8 \\
            J113508.918$+$501132.600 & GNIRS & 7200 & 6.5851 & [\ion{C}{II}] & 6 & 20.41 $\pm$ 0.16 & $-$26.075 & 10 & 12.8 & 11.6 & 16.2 \\
            J105807.720$+$293041.703 & NIRES/MODS & 3600/28800 & 6.5846 & [\ion{C}{II}] & 22 & 20.56 $\pm$ 0.05$^{\rm{c}}$ & $-$26.039 & 9 & 5 & 4.9 & 5.7 \\
            J092120.560$+$000722.900 & GNIRS & 9600 & 6.5646 & [\ion{C}{II}] & 6 & 21.11 $\pm$ 0.11 & $-$25.319 & 19 & 9 & 8.1 & 12 \\
            J022601.870$+$030259.280 & X-Shooter (NIR/VIS) & (14400/23520) & 6.5405 & [\ion{C}{II}] & 30 & 19.40 $\pm$ 0.09 & $-$27.192 & 18 & 26.5 & 30.6 & 11.7 \\
            J022426.540$-$471129.400 & X-Shooter (NIR/VIS) & (31200/33360) & 6.5222 & [\ion{C}{II}] & 6 & 19.73 $\pm$ 0.05 & $-$26.663 & 24 & 27.2 & 40.7 & 20 \\
            J043947.098$+$163415.819$^{\rm{a,g}}$ & X-Shooter (NIR/VIS) & (54720/53656) & 6.5192 & [\ion{C}{II}] & 6 & 17.47 $\pm$ 0.02 & $-$28.822 & 25 & 207.7 & 285.8 & 149.5 \\
            J111033.960$-$132945.600 & X-Shooter (NIR/VIS) & (34800/29520) & 6.5148 & [\ion{C}{II}] & 16 & 20.59 $\pm$ 0.18 & $-$26.079 & 18 & 4.8 & 7.6 & 3.7 \\
		\hline
	\end{tabular}
        \begin{tablenotes}
        \item Ref: 1 - \cite{Wang2021a}; 2 - \cite{Venemans2017}; 3 - \cite{Banados2018}; 4 - \cite{Yang2020a}; 5 - \cite{Mortlock2011}; 6 - \cite{Wang2021b}; 7 - \cite{Wang2018}; 8 - \cite{Yang2019}; 9 - \cite{Yang2021}; 10 - \cite{Wang2019}; 11 - \cite{Venemans2016}; 12 - \cite{Venemans2013}; 13 - \cite{Reed2019}; 14 - \cite{Banados2021}; 15 - \cite{Matsuoka2022}; 16 - \cite{Decarli2018}; 17 - \cite{Wang2017}; 18 - \cite{Venemans2015}; 19 - \cite{Matsuoka2018}; 20 - \cite{Mazzucchelli2017}; 21 - \cite{Tang2017}; 22 - \cite{Wang2024}; 23 - \cite{Banados2015}; 24 - \cite{Reed2017}; 25 - \cite{Fan2019}; 26 - \cite{Matsuoka2019b}; 27 - \cite{Banados2025}; 28 - \cite{Belladitta2025}; 29 - Bouwens et al. in preparation; 30 - \cite{Venemans2020}; 31 - \cite{Yang2023}.
        \item[a, b, c, d, e, f, g] Respectively: BAL quasar, $\rm{Y_{AB}}$ magnitude, $\rm{K_{p,AB}}$ magnitude, $M_{1450}$ from \cite{Matsuoka2019b}, unpublished quasar, $M_{1450}$ calculated as described in Appendix \ref{app:bal}, lensed quasar.
        \end{tablenotes}
        \end{threeparttable}
    \end{table}
\end{landscape}

\begin{table}
    \caption{Remaining \textit{published} quasars at $z>6.50$ that are not included in the sample. We report their short names, redshift, and the main reason why they are excluded from the analysis.}
    \centering
    \begin{tabular}{lc|c}
    \hline
     Quasar & $z$ & Reason for exclusion\\
    \hline
    J2356+0017 & 7.012 & Faint\\
    J1317+0017 & 6.94 & Faint\\
    J1349+0156 & 6.94 & Faint\\
    J1609+5328 & 6.92 & Faint\\
    J2210+0304 & 6.88 & Faint\\
    J0214+0232 & 6.829 & Faint\\
    J0112+0110 & 6.82 & Faint\\
    J1429$-$0104 & 6.796 & Faint\\
    J1310$-$0050 & 6.796 & Faint\\
    J1612+5559 & 6.78 & Faint\\
    J0244$-$5008 & 6.7306 & No echelle data\\
    J0229$-$0808 & 6.7249 & No echelle data\\
    J0213$-$0626 & 6.7214 & Faint\\
    J1344+0128 & 6.72 & Faint\\
    J1338+0018 & 6.701 & Faint\\
    J1205+0000 & 6.699 & Faint\\
    J0001+0000 & 6.69 & Faint\\
    J1231+0052 & 6.69 & Faint\\
    J1130+0450 & 6.68 & Faint\\
    J2252+0402 & 6.67 & Faint\\
    J0910+0056 & 6.65 & Faint\\
    J1035+0324 & 6.63 & Faint\\
    J1023+0044 & 6.63 & Faint\\
    J1450$-$0144 & 6.628 & Faint\\
    J0919+0512 & 6.62 & Faint\\
    J1732+6531 & 6.6 & No echelle data\\
    J1114+0215 & 6.55 & Faint\\
    J1440+0019 & 6.549 & Faint\\
    J0525$-$2406 & 6.5397 & No echelle data\\
    J1545+4232 & 6.5 & Faint\\
    \hline
    \end{tabular}
    \label{tab:excluded}
\end{table}

\subsection{Instruments}\label{subsec:instr}
We obtain visible (VIS) and NIR spectroscopy of our sample from the data archives of three main echelle spectrographs: Gemini/GNIRS (Gemini Near-Infrared Spectrograph; \citealt{Elias2006a,Elias2006b}), Keck/NIRES (Near-Infrared Echellette Spectrometer; \citealt{Wilson2004}), and VLT/X-Shooter \citep{Vernet2011}, both VIS and NIR arms, except for J1342$+$0928, whose spectrum is entirely covered by the NIR arm. For some quasars, part of the spectrum redward of Ly$\alpha$ is in the optical, and hence not covered by the NIR arm, so we have complemented them using observations from additional long slit instruments, such as Gemini/GMOS (Gemini Multi-Object Spectrographs; \citealt{Hook2004}), Keck/DEIMOS (DEep Imaging Multi-Object Spectrograph; \citealt{Faber2003}), Keck/LRIS (Low-Resolution Imaging Spectrometer; \citealt{Oke1995, Rockosi2010}) and LBT/MODS (Multi-Object Double Spectrographs; \citealt{Pogge2010}). A complete list of the instruments used to observe each quasar is available in Table \ref{tab:sample}.

\begin{table*}
    \caption{Properties of the instruments used to collect the data. The black line separates the NIR from the VIS arm.}
    \label{tab:instr_prop}
    \begin{threeparttable}
    \begin{tabular}{lcccccc}
	\hline
	Instrument & Dither sequence & $\Delta \lambda$ ($\rm{\mu m}$) & Slit width ($^{\prime\prime}$) & $R$ & FWHM (km s$^{-1}$) & $dv_{\rm{pix}}$ (km s$^{-1}$)$^{\mathrm{a}}$ \\
	\hline
        GNIRS & ABBA & $0.81-2.52$ & 0.68/1.00 & $\simeq 1200/ \simeq 1000$ & $\simeq 280$ & $\simeq 86$ \\
        NIRES & ABBA & $0.94-2.47$ & 0.55 & $\simeq 2700$ & $\simeq 111$ & $\simeq 38$ \\
        X-Shooter NIR & varies & $1.02-2.48$ & 0.6 & $\simeq 8100$ & $\simeq 37$ & $\simeq 13$ \\ \hline
        X-Shooter VIS & - & $0.55-1.02$ & 0.9 & $\simeq 8900$ & $\simeq 34$ & $\simeq 11$ \\
        GMOS & - & $0.65-1.15$ & 1.00 & $\simeq 3200$ & $\simeq 94$ & $\simeq 51$ \\
        DEIMOS & - & $0.65-1.00$ & 1.00 & $\simeq 7900$ & $\simeq 38$ & $\simeq 17$ \\
        LRIS & - & $0.70-1.03$ & 1.00 & $\simeq 1700$ & $\simeq 176$ & $\simeq 56$ \\
        MODS & - & $0.42-1.12$ & 1.00/1.2 & $\simeq 2700/ \simeq 2300$ & $\simeq 111$ & $\simeq 33$ \\
	\hline
    \end{tabular}
    \begin{tablenotes}
        \item[a] The pixel sizes reported are those of each individual spectrograph, while the ones used to re-bin the spectra during the co-addition procedure are reported in Section \ref{subsec:coadd}, and chosen to be coarser than the coarsest $dv_{\rm{pix}}$ between the involved instruments.
    \end{tablenotes}
    \end{threeparttable}
\end{table*}

\subsubsection{Dither sequences}
All GNIRS and NIRES observations were executed following an ABBA dither sequence except for the last two frames of J0313$-$1806, which are AABB. The X-Shooter observations were acquired with different dither sequences (i.e.: ABBA, ABAB, AABB), but because of the long average exposure time per frame ($\simeq 600-1800$ s) and the consequent change of the sky conditions, they were treated as ABCD during the data reduction instead of performing usual image differencing. The only X-Shooter frames where a dither sequence of ABBA was used while performing the reduction are those for J1110$-$1329 because of the short exposure time of $300$ s. This applies mostly for the NIR arm since VIS data are usually not reduced doing image differencing, but using a sky model.

\subsubsection{Wavelength coverage, Slit widths, $R$, FWHM, and $dv_{\rm{pix}}$}
The GNIRS data provide full wavelength coverage of the YJHK bands ($0.81-2.52$ $\mu$m). Slit widths of 0.68$^{\prime\prime}$ or 1.00$^{\prime\prime}$ were used, with mean resolution of $R \simeq 1200-1000$ or $\rm{FWHM} \simeq 280$ km s$^{-1}$ and pixel size of $dv_{\rm{pix}}\simeq 86$ km s$^{-1}$.
The NIRES observations have a coverage of $0.94-2.47$ $\mu$m through a fixed 0.55$^{\prime\prime}$ slit, where the mean resolution is $R \simeq 2700$ or $\rm{FWHM} \simeq 111$ km s$^{-1}$ and pixel size of $dv_{\rm{pix}}\simeq 38$ km s$^{-1}$.
The X-Shooter data cover the wavelength range $0.55-1.02$ $\mu$m in the VIS arm and $1.02-2.48$ $\mu$m in the NIR arm. The slit width varies among programs but is typically 0.9$^{\prime\prime}$ in the VIS and 0.6$^{\prime\prime}$ in the NIR arm. The mean resolution in VIS is $R \simeq 8900$ or $\rm{FWHM} \simeq 34$ km s$^{-1}$ and pixel size $dv_{\rm{pix}}\simeq 11$ km s$^{-1}$; while in NIR, $R \simeq 8100$ or $\rm{FWHM} \simeq 37$ km s$^{-1}$ and $dv_{\rm{pix}}\simeq 13$ km s$^{-1}$.
J0319$-$1008 is complemented with R400 grating GMOS-N observations, covering the wavelength range $0.65-1.15$ $\mu$m. The slit width is 1.00$^{\prime\prime}$, with a mean resolution of $R \simeq 3200$ or $\rm{FWHM} \simeq 94$ km s$^{-1}$, and pixel size of $dv_{\rm{pix}}\simeq 51$ km s$^{-1}$.
J0706$+$2921 is complemented with 830G grating DEIMOS observations, covering the wavelength range $0.65-1.00$ $\mu$m. The slit width is 1.00$^{\prime\prime}$, with a mean resolution of $R \simeq 7900$ or $\rm{FWHM} \simeq 38$ km s$^{-1}$, and pixel size of $dv_{\rm{pix}}\simeq 17$ km s$^{-1}$.
J0218$+$0007 is complemented using grating 600/10000 LRIS red observations, covering the wavelength range $0.70-1.03$ $\mu$m. The slit width is 1.00$^{\prime\prime}$, with a mean resolution of $R \simeq 1700$ or $\rm{FWHM} \simeq 176$ km s$^{-1}$, and pixel size of $dv_{\rm{pix}}\simeq 56$ km s$^{-1}$.
J0411$-$0907, J1917$+$5003, and J1058$+$2930 are complemented by red grating MODS observations (both MODS1 and MODS2), covering the wavelength range $0.42-1.12$ $\mu$m. The slit width is 1.00$^{\prime\prime}$ and 1.2$^{\prime\prime}$, with a mean resolution of $R \simeq 2700-2300$ or $\rm{FWHM} \simeq 111$ km s$^{-1}$, and pixel size of $dv_{\rm{pix}}\simeq 33$ km s$^{-1}$. All of these properties of the instruments are summarised in Table \ref{tab:instr_prop}.

\subsection{Data Reduction}\label{sec:dataredux}
All spectra are reduced with the open-source Python-based Spectroscopic Data Reduction Pipeline \texttt{PypeIt}\footnote{\url{https://github.com/pypeit/PypeIt}}, using versions between 1.7.1 and 1.14.1 \citep{Prochaska2020a, Prochaska2020}. The pipeline performs image processing, including gain correction, bias subtraction, dark subtraction, and flat fielding. It uses supplied flat-field images to automatically trace the echelle orders and correct for the detector illumination. Construction of the wavelength solutions and the wavelength tilt models are based on either arc (for VIS instruments) or science frames (i.e. using sky OH lines, for NIR spectrographs). Cosmic rays are removed with the L. A. COSMIC algorithm \citep{vanDokkum2001}. The sky subtraction is based on the standard A–B mode and a B-spline fitting procedure that is performed to further clean up the sky line residuals following \cite{Bochanski2009}. Optimal extraction \citep{Horne1986} is performed to generate 1D science spectra. 
We clarify that \texttt{PypeIt} uses a nearest grid point interpolation algorithm for the extraction and co-addition of spectra to avoid correlations between pixels.
The extracted spectra are flux-calibrated with sensitivity functions derived from the observations of spectroscopic standard stars. All flux-calibrated 1D spectra of each quasar are then co-added to achieve higher SNR and corrected for telluric absorption using \texttt{PypeIt}.
In particular, during the co-addition procedure, both spectra and noise vectors are iteratively rescaled to a preliminary mean spectrum, then they are optimally weighted and combined. This procedure is continued in a loop with the rescaling factors converging to unity as the procedure is iterated. To conclude, a telluric model is fit to correct the absorbed science spectrum up to a best-fit PCA model \citep{Davies2018b} of said spectrum. The telluric model is based on telluric model grids produced from the Line-By-Line Radiative Transfer Model (LBLRTM4; \citealt{Clough2005,Gullikson2014}).

All the \texttt{PypeIt} files to reproduce the reduction are publicly available in a GitHub repository\footnote{\url{https://github.com/enigma-igm/onorato24_hiz_qsos}}.

\subsection{Co-add of spectra from different instruments (or arms)}\label{subsec:coadd}
As reported in Table \ref{tab:sample} and in Section \ref{subsec:instr}, some quasars are observed with more than one spectrograph and/or in more arms. Here we describe how we treat these spectra after the flux calibration with sensitivity function since they have different resolutions, wavelength grids, and pixel sizes.
\begin{itemize}
    \item X-Shooter VIS - NIR: two different approaches were used depending on the stage of development of \texttt{PypeIt}. We co-add the majority of the quasars with versions between 1.8.2 and 1.11.1, using the echelle combspec \texttt{PypeIt} routine, getting a unique spectrum in a wavelength grid of [5410, 24770] {\angstrom} and pixel size of $dv_{\rm{pix}}\simeq 13$ km s$^{-1}$. Few quasars were reduced using recent \texttt{PypeIt} versions (after 1.12.3) which now support the option of editing a unique file (\texttt{.coadd1d}) to combine the 1D spectra from multiple exposures of the same object, in case we work with the same type of spectrograph (i.e. echelle), containing both arms together. The final spectrum has the same characteristics as those obtained with the first method.
    \item GNIRS - NIRES: we co-add them using a single \texttt{coadd1d} file. J0313$-$1806 and J1007$+$2115 are co-added onto a common grid covering [9410, 24690] {\angstrom} with a pixel size of $dv_{\rm{pix}}\simeq 90$ km s$^{-1}$. We co-add J0910$-$0414 and J2102$-$1458 by requiring a pixel size of $dv_{\rm{pix}}\simeq 90$ km s$^{-1}$, but without constraints on the wavelength range. We get a final spectrum that covers [8240$-$25200] {\angstrom} and has $dv_{\rm{pix}}\simeq 90$ km s$^{-1}$.
    \item GNIRS - X-Shooter (VIS - NIR): we co-add them using a unique \texttt{coadd1d} file for all the GNIRS, X-Shooter VIS, and NIR frames. The final spectra cover [5410, 25200] {\angstrom} with a pixel size of $dv_{\rm{pix}}\simeq 90$ km s$^{-1}$.
    \item NIRES - GMOS: we co-add all the echelle - long slit spectra using the multi combspec \texttt{PypeIt} routine, which works with the final 1D spectrum in both instruments, getting an ultimate spectrum with wavelength coverage of [8000, 24700] {\angstrom} and pixel size of $dv_{\rm{pix}}\simeq 55$ km s$^{-1}$.
    \item NIRES - DEIMOS: using multi combspec as described above, we get a final spectrum covering [8000, 24700] {\angstrom} with pixel size of $dv_{\rm{pix}}\simeq 40$ km s$^{-1}$.
    \item NIRES - LRIS: using multi combspec as described above, we get a final spectrum covering [8000, 24700] {\angstrom} with pixel size of $dv_{\rm{pix}}\simeq 60$ km s$^{-1}$.
    \item NIRES - MODS: using multi combspec as described above, we get a final spectrum covering [8000, 24700] {\angstrom} with pixel size of $dv_{\rm{pix}}\simeq 40$ km s$^{-1}$.
\end{itemize}

In the end, when echelle frames are co-added together, the telluric correction described above (at the end of Section \ref{sec:dataredux}) is applied to the final stacked spectrum; when echelle - long slit are co-added, the individual spectra of both instruments are already telluric corrected, and no further correction is required.
In principle, telluric correction should be applied before heliocentric correction, since telluric absorption occurs in the Earth's atmospheric frame, not the heliocentric vacuum frame. However, for many of the faint quasars in this work, individual exposures have low SNR, making accurate telluric correction difficult using the model-based approach in \texttt{PypeIt}. Moreover, our data were often taken on different nights, and aligning exposures in the Earth's rest-frame would shift the quasar spectrum, complicating outlier rejection. We therefore apply telluric correction to the co-added, heliocentric-frame spectra as a compromise, accepting minor line broadening to ensure sufficient SNR for accurate modelling and outlier removal. Also, this broadening effect is more relevant for bright sources, and less critical for the faint quasars in our sample.

\begin{figure}
    \includegraphics[width=\linewidth]{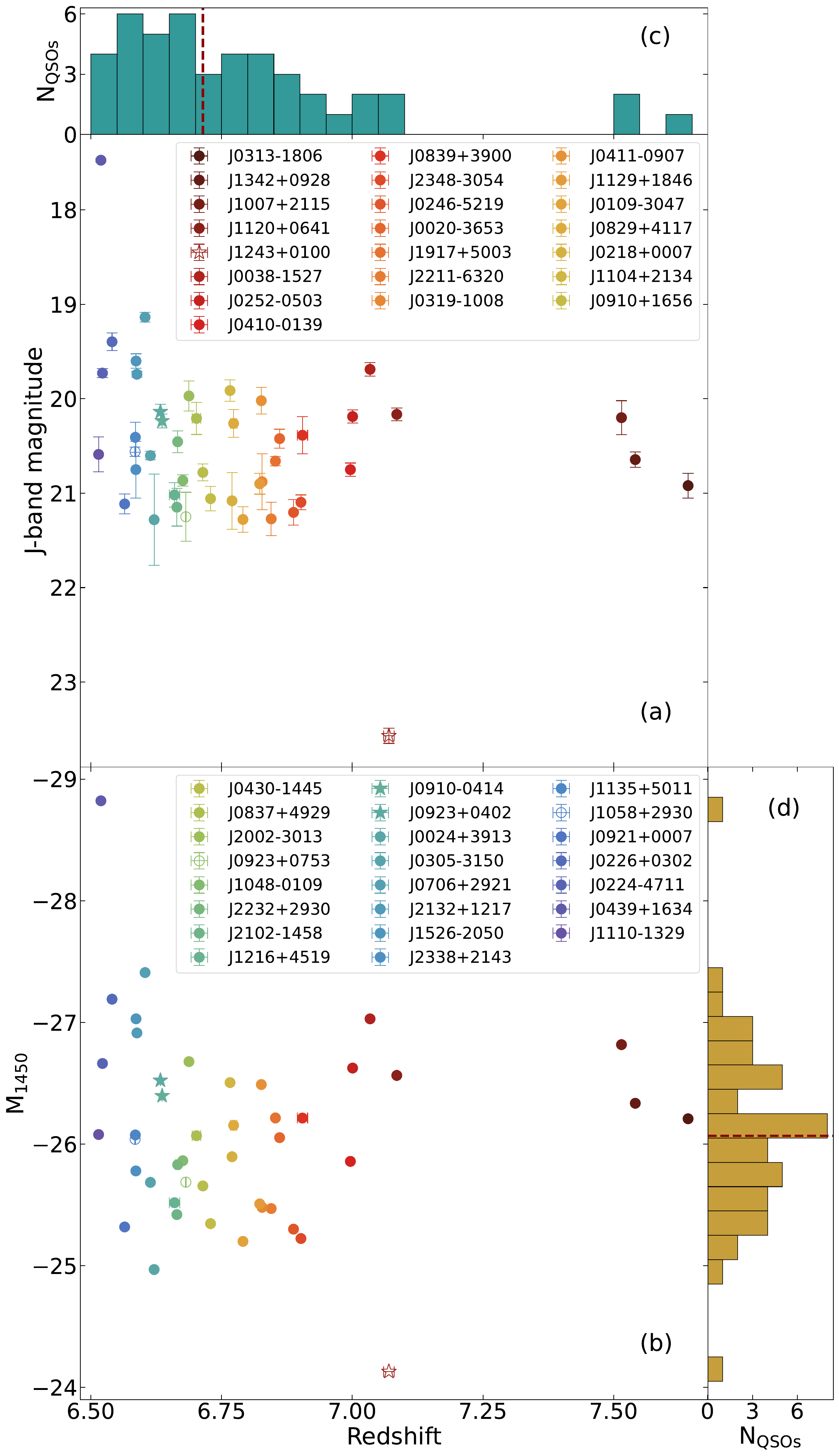}
    \caption{\textit{Panel (a)}: Distribution of J-band photometry used to scale the spectra as a function of $z$ for all the $45$ quasars in this sample. The legend is unique and split between the two plots, showing the sources color-coded with decreasing $z$. The full symbols mark the quasars for which J-band photometry is available, while the open symbols mark those for which it is not (we report Y-band photometry for J1243$+$0100 and J0923$+$0753, and $\rm{K_{p}}$-band photometry for J1058$+$2930). The error bars show the uncertainties on both $z$ and the photometric measurements. \textit{Panel (b)}: Distribution of $M_{1450}$ as a function of $z$ for all the $45$ quasars in this sample. The error bars show the uncertainties on $z$. The circles mark the $M_{1450}$ computed from the spectrum scaled with J, Y, or $\rm{K_{p}}$-band photometry. The stars mark the quasars for which $M_{1450}$ from spectroscopy is not reliable because of the low SNR of the spectrum or appearance of BAL features. In the case of J1243$+$0100, $M_{1450}$ is from the discovery paper (\citealt{Matsuoka2019b}); while for J0910$-$0414 and J0923$+$0402, we follow the method described in Appendix \ref{app:bal}. \textit{Panel (c)}: Histogram of the redshift distribution of the sample, with bins of size 0.05. The dashed red line represents the median redshift ($z_{\rm{median}}=6.71$). \textit{Panel (d)}: Histogram of the $M_{1450}$ distribution of the sample, with bins of size 0.2. The dashed red line represents the median $M_{1450}$ ($M_{1450,\rm{median}} \simeq -26.1$).}
    \label{fig:M1450}
\end{figure}

\begin{table*}
    \caption{Example of the internal structure of the \texttt{FITS} files, at a random row, for two of the spectra in this sample. The first column is the wavelengths in the observed frame in units of {\angstrom}. The second column is a new wavelength grid evaluated at the centers of the wavelength bins, also in units of {\angstrom} (this is the quantity we usually put on the $x$-axis when plotting a spectrum). The third column is the flux array in units of $10^{-17}$ erg s$^{-1}$cm$^{-2}${\angstrom}$^{-1}$. The fourth column is the inverse variance array in units of ($10^{-17}$ erg s$^{-1}$cm$^{-2}${\angstrom}$^{-1}$)$^{-2}$. "sigma" is a column introduced only in the most recent \texttt{PypeIt} versions and represents the noise vector (if it is not present, it is always possible to determine from ivar$^{-1/2}$). The sixth column is a good pixel mask for the spectrum, and the seventh one is the telluric model. The "obj model" column is not present in the spectra reduced with the most recent \texttt{PypeIt} versions and represents the object model used for the telluric fitting. The last two columns are created after the flux scaling procedure and represent the new flux and ivar corrected with the $X$-band photometry of the quasar, where $X$ could be J, Y, or $\rm{K_{p}}$ depending on the available photometric information, in the same units as before.}
    \centering
    \begin{tabular}{l|cccccccccc}
    \hline
      & wave & wave grid mid & flux & ivar & sigma & mask & telluric & obj model & flux scaled $X$ & ivar scaled $X$\\
    \hline
    \textbf{J1342$+$0928} & 10289.82613 & 10289.83458 & 0.31196 & 18.34511 & - & 1 & 0.99 & 0.61569 & 0.36666 & 13.27947 \\ \hline
    \textbf{J1058$+$2930} & 10240.48127 & 10240.45124 & 1.56780 & 2.58264 & 0.62226 & 1 & 1.0 & - & 0.54980 & 21.00057 \\
    \hline
    \end{tabular}
    \label{tab:spectra}
\end{table*}

\section{Quasar sample Properties}\label{sec:qsoprop}
\subsection{Redshift}\label{subsec:z_err}
In this paper, we account for uncertainties in the systemic redshift of every quasar depending on the emission line used to determine its redshift. Systemic redshifts are challenging quantities to determine because of various factors, like the broad widths of emission lines, Gunn-Peterson absorption \citep{Gunn1965}, and offsets between different ionization lines (\citealt{Gaskell1982}; \citealt{Tytler1992}; \citealt{VandenBerk2001}; \citealt{Richards2002}; \citealt{Shen2016}). Also, most quasars show winds and strong internal motions, which displace many of the emission lines far from the systemic redshift of the host galaxy.
Thus, following \cite{Eilers2017}, we decide to assign a redshift error of $\Delta v = 100$ km s$^{-1}$ for the most precise determinations of the location of the quasar, such as those using emission lines from the atomic gas reservoir of the host galaxy itself ([\ion{C}{II}] lines). However, we emphasize that obtaining this measurement is not straightforward: the typical velocity dispersion ($\sigma_v$) observed within quasar host galaxies is around $300$ km s$^{-1}$ for detections of at least SNR $> 5$. The statistical error on the mean redshift is approximately $\sigma_v/\rm{SNR}\approx 60$ km s$^{-1}$. Thus, our adopted uncertainty of $100$ km s$^{-1}$ is just a conservative estimate. This assumption is based on the expectation that the quasar resides within the gravitational potential traced by the cool interstellar medium. For quasars with a redshift measurement from low-ionization broad emission lines, such as \ion{Mg}{II}, we assume a redshift error of $\Delta v = 390$ km s$^{-1}$, to account for the dispersion between the redshift of the \ion{Mg}{II} line and the redshift of the host galaxy (e.g., \citealt{Mazzucchelli2017, Schindler2020}). This value is supported by previous literature and further confirmed by our own analysis on a sub-sample of our sources with both \ion{Mg}{II} and [\ion{C}{II}] redshifts, for which we find a median velocity shift of $\Delta v \simeq -388$ km s$^{-1}$, in excellent agreement with \cite{Schindler2020}.
Table \ref{tab:sample} reports redshifts, methods, and references for every quasar.

These $\Delta v$ will be adopted as uncertainties on the redshift (and thus on the position of the Ly$\alpha$ line) to achieve our next scientific goal of studying the proximity zones of the quasars in this sample \citep{Onorato2025}. From the literature ($z_{\rm{Ref}}$ in Table \ref{tab:sample}), we also get values of $\Delta z$ for every quasar, but we do not report them here. They are available online at the link at the end of Section \ref{sec:dataredux}, and we use them as uncertainties on $z$ in Figure \ref{fig:M1450}.

\subsection{Absolute flux calibration}\label{subsec:fluxscale}
The flux calibration of the spectra obtained by the data reduction is relative: every spectrum is corrected using a spectroscopic standard star, but the flux values at the different wavelengths may differ from the true flux. This could be due to slit losses and non-photometric sky conditions at the time of observations.
Tests of absolute flux calibration have been performed in this analysis, tying the spectra to the photometric data in the Y, J, H, and K bands, where all or part of them are available. However, since there is not a good match in all the photometric bands at once, a definitive method is still an object of discussion. Also, we point out that the photometry is not taken together with the spectra, and this could introduce uncertainty in the absolute flux calibration, as some of these quasars might have varied between the spectroscopic and photometric observations.

For this paper, the reduced spectrum of each quasar is scaled using its J-band magnitude, or the Y and $\rm{K_{p}}$ band ones if the J-band is not available (as in the case of J1243$+$0100, J0923$+$0753, and J1058$+$2930), all in the AB system; an example for all bands is shown in Figure \ref{fig:scaled} and the approach is described in Appendix \ref{app:fluxscale}. 
We also evaluate the consistency of this scaling by comparing synthetic photometry from the flux-scaled spectra to available Y, H, and K-band magnitudes. We find that the typical discrepancy is around $10\%$, with a maximum deviation of $\sim 40\%$ (see Appendix \ref{app:fluxscale}).
These NIR data come from different archival surveys: the UKIRT Infrared Deep Sky Survey (UKIDSS; \citealt{Lawrence2007}), the UKIRT Hemisphere Survey (UHS; \citealt{Dye2018}),
and the VISTA Hemisphere Survey (VHS; \citealt{McMahon2013}). The Y-band photometry of J1243$+$0100 comes from the Hyper Suprime-Cam (HSC) Subaru Strategic Program (SSP) survey \citep{Aihara2018}, as reported in the discovery paper \citep{Matsuoka2019b}. The J-band photometry of J0410$-$0139, J0430$-$1445, and J2132$+$1217 is taken from SofI (Son of ISAAC; \citealt{Moorwood1998}), at the New Technology Telescope (NTT) in La Silla, and that of J1129$+$1846 and J1917$+$5003 comes from NOTCam\footnote{Both SofI and NOTCam are dedicated follow up.}, at the Nordic Optical Telescope (NOT). The $\rm{K_{p}}$-band photometry of J1058$+$2930 is gotten from the acquisition image taken with NIRES, whose guider camera uses a Mauna Kea K$^{\prime}$ ("K prime") filter\footnote{\url{https://www2.keck.hawaii.edu/inst/nires/guider.html}}. Since the transmission curve of this filter is not available, we use the one from Keck/NIRC2 $\rm{K_{p}}$ to perform the flux scaling (see the bottom panel of Figure \ref{fig:scaled}), as it is very similar to the Mauna Kea K$^{\prime}$.
Most of these magnitudes are collected from \cite{Ross2020} and available online. All J-band (or Y and $\rm{K_{p}}$) magnitudes are displayed in panel (a) of Figure \ref{fig:M1450} as a function of redshift $z$ for each quasar, and the final flux-scaled spectra are plotted as a function of wavelength in Figure \ref{fig:spectra}.

A detailed list of all the filters that were used to acquire every magnitude is available at the link given at the end of Section \ref{sec:dataredux}, while the final flux calibrated spectra are available online as explained in the Data Availability Section. An example of the format of the \texttt{FITS} files, selecting a random row, for two spectra reduced using different \texttt{PypeIt} versions is shown in Table \ref{tab:spectra}.

\subsection{Absolute magnitudes at 1450 {\angstrom}}\label{subsec:M1450}
We calculate $M_{1450}$ using the flux-scaled spectra\footnote{As we scale almost all the spectra using the J-band, which is the photometric band closest to $1450 \cdot (1 + z)$ {\angstrom}, any possible uncertainties in the spectral shape should not result in a significant source of error on $M_{1450}$.} of the quasars themselves. From every spectrum in the observed frame and knowing the redshift of each quasar, we compute the rest-frame wavelength. 

We determine the apparent AB magnitude at $1450$ {\angstrom} from the median flux between $1445$ and $1455$ {\angstrom}, converted to Jansky ($f_{\nu,1450}$), using the Pogson law: $m_{1450} = -2.5 \cdot \log_{10}(f_{\nu,1450}) + 8.9$.
We continue calculating the luminosity per unit frequency ($L_{\nu}$) at $1450$ {\angstrom} with Equation \ref{eq:luminosity}, where $d_{L}$ is the luminosity distance to the object at a redshift $z$:
\begin{equation}\label{eq:luminosity}
L_{\nu} = \frac{4\pi {d_{L}}^{2} \cdot f_{\nu,1450}}{1 + z} 
\end{equation}
We find the absolute AB magnitude at $1450$ {\angstrom} using Equation \ref{eq:M1450} from the luminosity per unit frequency $L_{\nu}$, where $d_0$ is the reference distance ($10$ pc), and $3631$ Jy is the zero-point flux density in the AB system:
\begin{equation}\label{eq:M1450}
M_{1450} = -2.5 \cdot \log_{10} \left( \frac{L_{\nu}}{4 \pi {d_{0}}^{2} \cdot 3631 \text{Jy}} \right)
\end{equation}
The results from this method are compared with other values from the literature, showing good agreement. For this reason, we trust the estimates we obtain for all quasars, with only three exceptions. In the case of J1243$+$0100, the spectrum is of poor quality as the source is faint and has a low SNR (see the values at the last three columns of Table \ref{tab:sample}). As a consequence, this makes the result difficult to trust, so we adopt the $M_{1450}$ from \cite{Matsuoka2019b}, in which they measured it from the best-fit power-law continuum.
In the case of J0910$-$0414 and J0923$+$0402, the BAL features visible in their spectra create a bias in the measurement of $M_{1450}$. For this reason, we correct the estimate by matching these spectra with a "reference spectrum" with a trustworthy continuum shape, such as the composite spectrum created from the sample in this paper (shown in Figure \ref{fig:composite}). The match is performed by eye and shown in Figure \ref{fig:matchbal}, with a detailed description of the method followed reported in Appendix \ref{app:bal}.

We show the absolute magnitude values $M_{1450}$ as a function of redshift $z$ for our sample in panel (b) of Figure \ref{fig:M1450}, with its histogram in panel (d).

\begin{figure*}
    \centering
    \includegraphics[width=16.5cm]{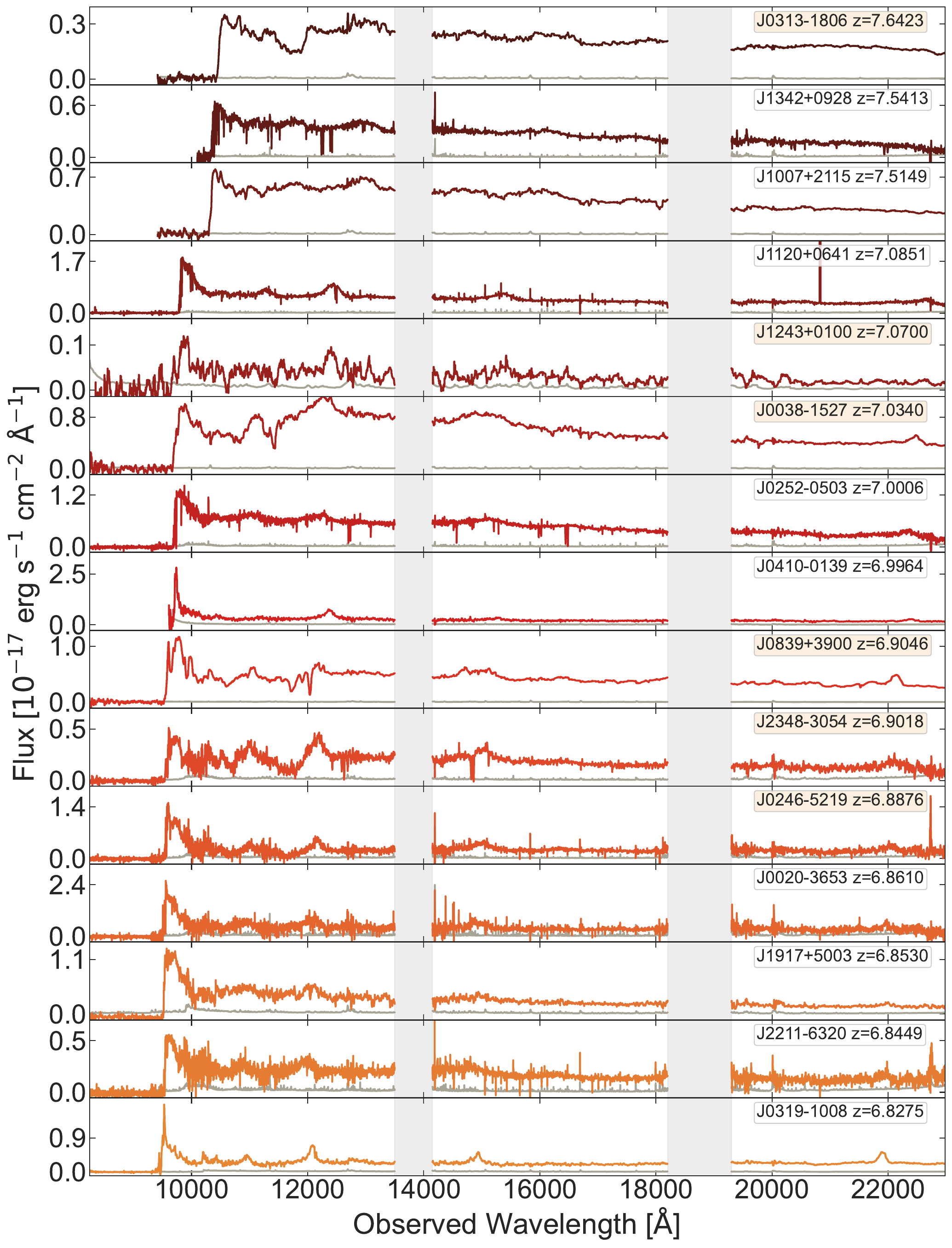}
     \caption{Spectra of all the $45$ quasars in this sample, sorted by decreasing $z$ (as reported in Table \ref{tab:sample}). Every subplot shows the spectrum (color-coded with $z$) and the associated error (grey) in the observed wavelength range [8245, 22980] {\angstrom}, the short name, and $z$ of the quasar. The light-grey bands cover [13500, 14150] {\angstrom}, and [18200, 19300] {\angstrom}, indicating the regions affected by strong telluric absorption, where the spectra are masked. All the spectra shown here are smoothed for display purposes. The box containing the name and the redshift of the quasars is colored in light beige to mark the BALs.}
     \label{fig:spectra}
\end{figure*}

\begin{figure*}
    \addtocounter{figure}{-1}
    \centering
    \includegraphics[width=17cm]{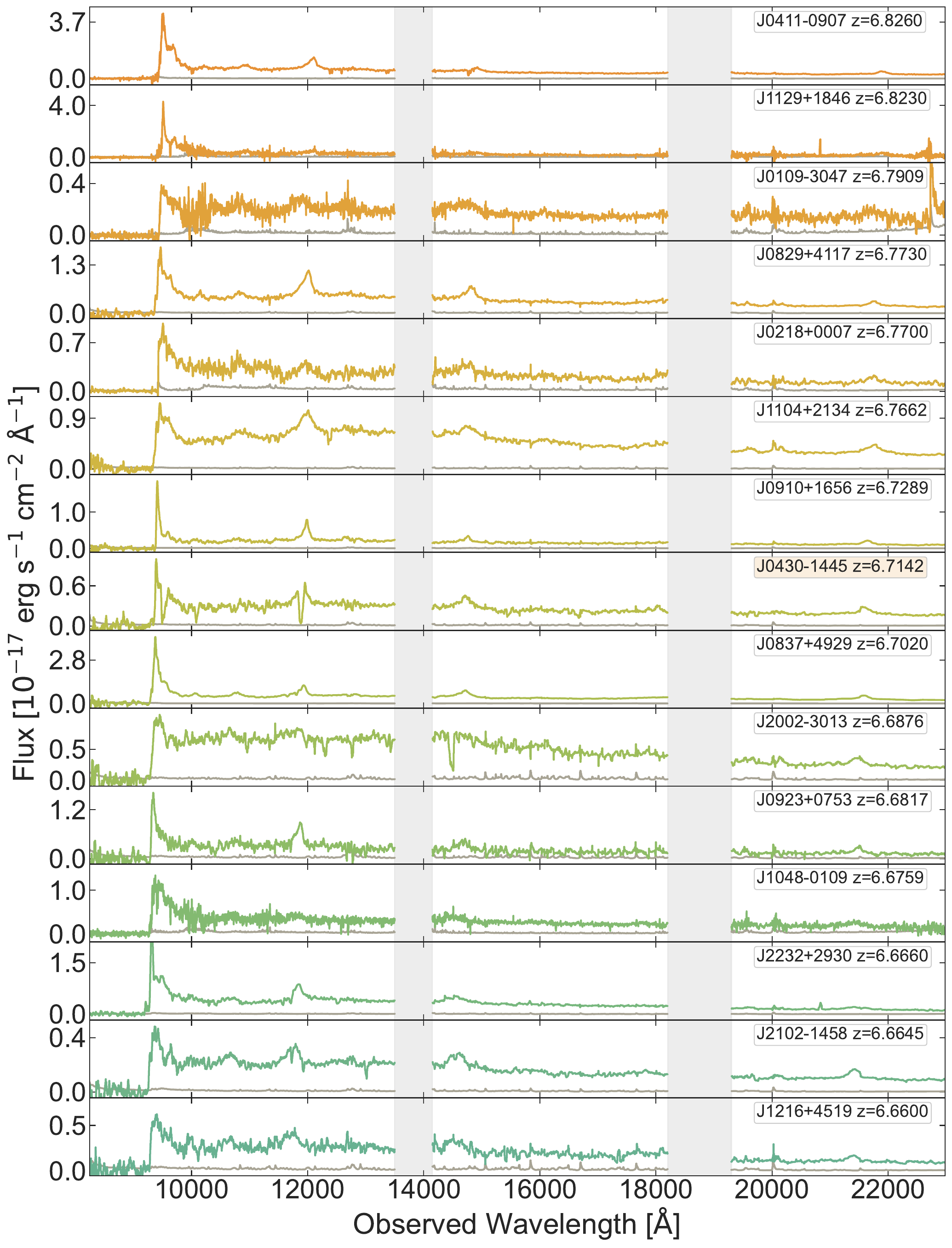}
     \caption{(Continued)}
\end{figure*}

\begin{figure*}
    \addtocounter{figure}{-1}
    \centering
    \includegraphics[width=17cm]{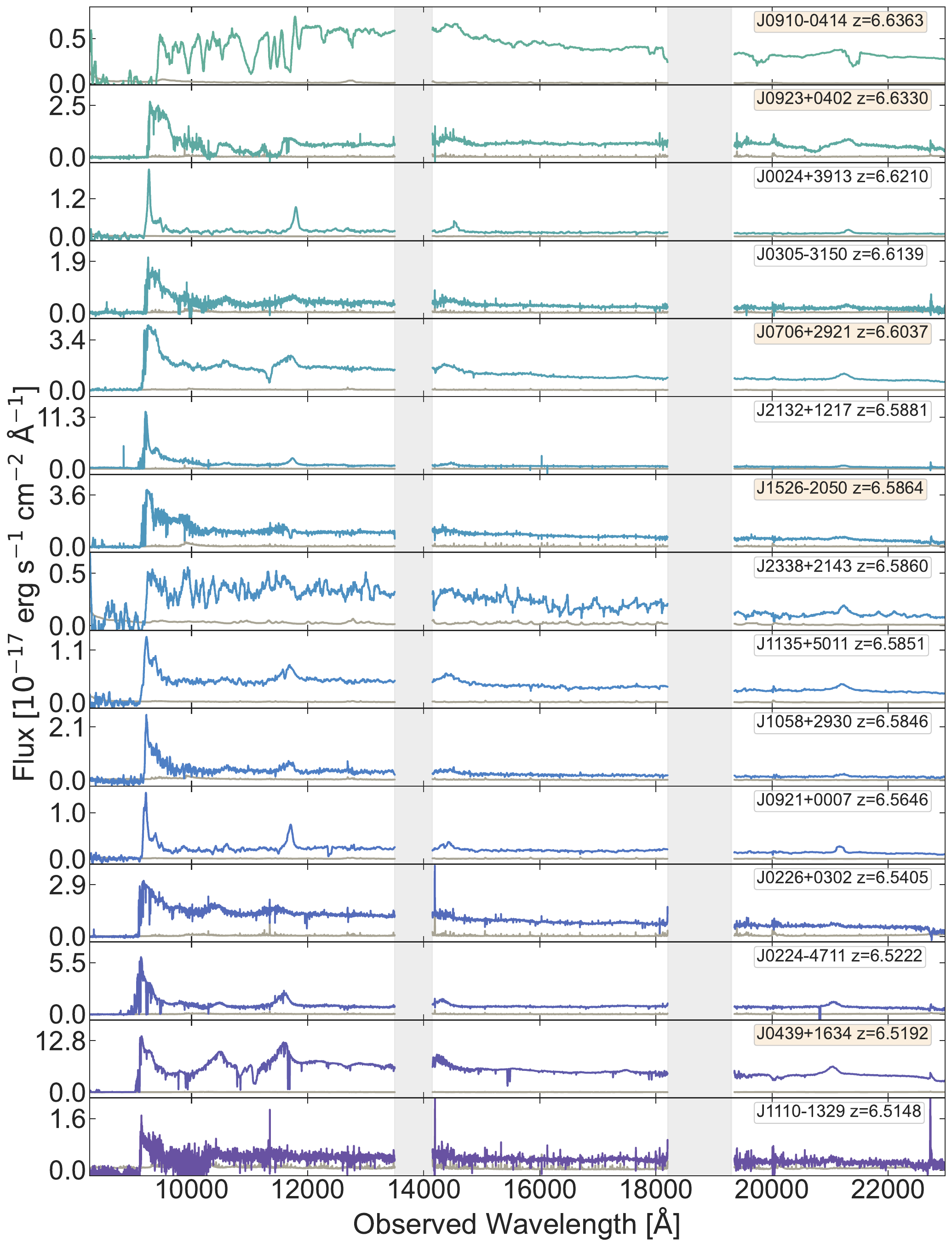}
     \caption{(Continued)}
\end{figure*}

\subsection{SNR of the spectra}\label{sec:snr}
We compute the SNR of the spectra in this sample considering three different wavelength ranges (the J, H, and K bands). We want to properly sample the spectra and avoid possible biases introduced by telluric absorption regions. For these reasons, we define the ranges in which we can compute the SNR in a very conservative manner, avoiding regions affected by absorption: $[11000, 13400]$, $[14500, 17950]$, and $[19650, 22400]$ {\angstrom}.

To make the calculation and have a comparison among the quality of the spectra that is as fair as possible, we follow these steps:
\begin{itemize}
    \item{We move the spectra to the rest-frame, dividing the wavelengths by $(1+z)$;}
    \item{We define a new wavelength grid: [1040, 3332] {\angstrom} in steps of $dv_{\rm{pix}}$=110 km s$^{-1}$ (set to be coarser than the coarsest pixel scale, in velocity, among all the instruments used to create the sample);}
    \item{We re-bin the spectra into the new wavelength grid;}
    \item{We finally compute the mean SNR in the three different wavelength ranges ($\langle \rm{SNR}_{\lambda} \rangle$, where $\lambda=$ J, H, K), also shifted to the rest-frame according to the redshift of the quasar considered}.
\end{itemize}

\begin{figure}
  \includegraphics[width=\linewidth]{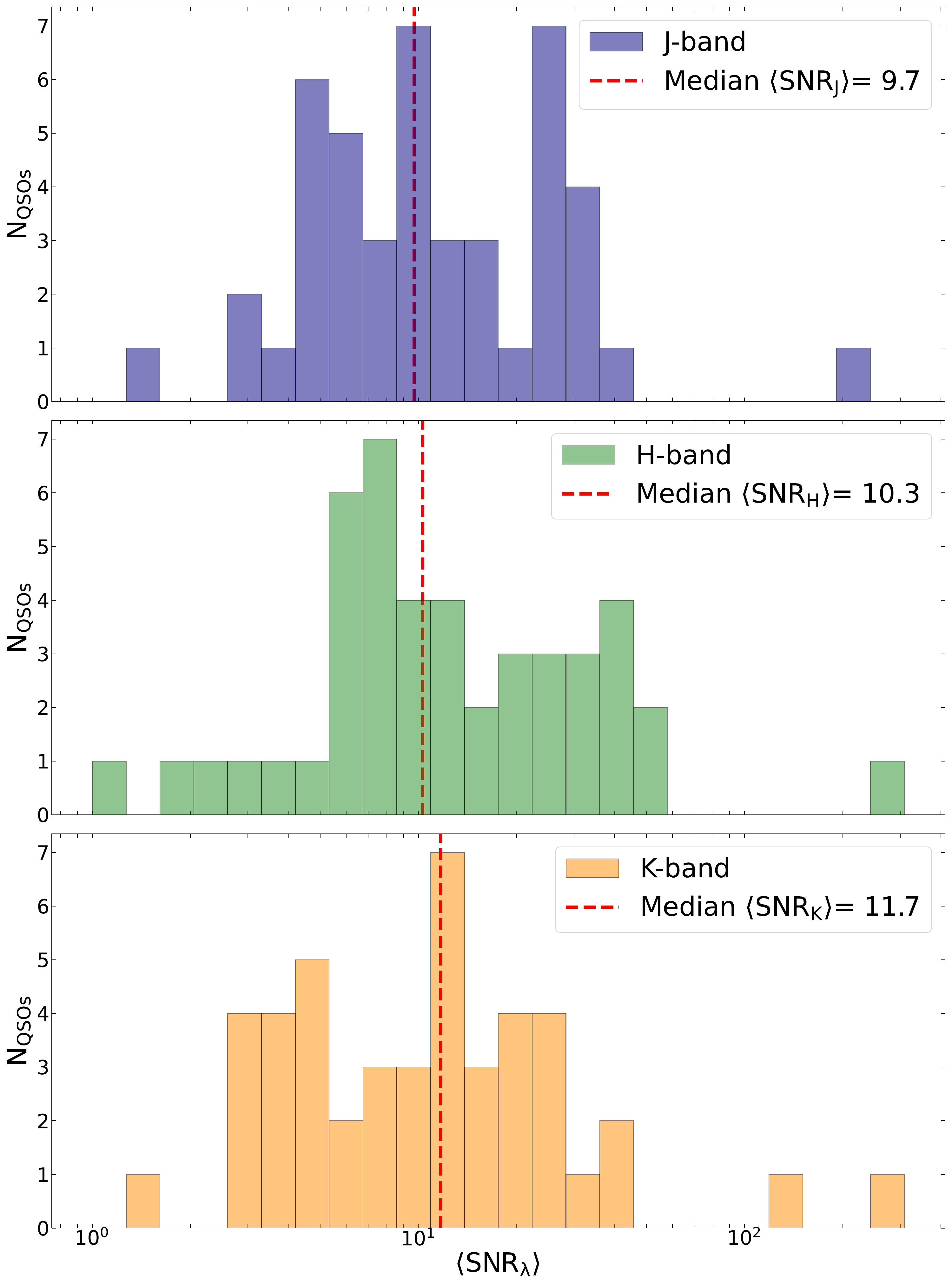}
 \caption{\textit{Top}: Histogram of the $\langle \rm{SNR}_{\lambda} \rangle$ of the quasars in this sample, reported in Table \ref{tab:sample} in the J-band. We generate 25 logarithmically spaced bins between the minimum and maximum of all the $\langle \rm{SNR}_{\lambda} \rangle$ values in the sample. The median $\langle \rm{SNR}_{\lambda} \rangle$ is shown as a red dashed vertical line. \textit{Middle}: Same, but for the H-band. \textit{Bottom}: Same, but for the K-band.}
 \label{fig:snr_hist}
\end{figure}

Two examples showing the method described above, considering the highest and lowest redshift quasar in the sample, are visible in Figure \ref{fig:snr}.
The results of this analysis are shown in the last three columns of Table \ref{tab:sample}, and in the histograms in Figure \ref{fig:snr_hist}, where we generate 25 logarithmically spaced bins between the minimum and maximum
of all the $\langle \rm{SNR}_{\lambda} \rangle$ values in the sample.
The majority of the spectra have $\langle \rm{SNR}_{J,H} \rangle$ in the range [5,10], and $\langle \rm{SNR}_{K} \rangle$ between $0-5$ and $10-15$. To quantify the overall data quality we compute the median $\langle \rm{SNR}_{\lambda} \rangle$ in the three wavelength ranges.
We have median $\langle \rm{SNR}_{J} \rangle=9.7$, median $\langle \rm{SNR}_{H} \rangle=10.3$, and median $\langle \rm{SNR}_{K} \rangle=11.7$ (reported as red dashed vertical lines in Figure \ref{fig:snr_hist}).

The different properties for every quasar are reported in Table \ref{tab:sample}, which lists: the name of the quasars, the instruments (and arms), the exposure time for each quasar in every instrument, their redshifts, the method adopted to determine the redshift, the reference for this measurement, the J-band photometry in the AB system, the magnitude ($M_{1450}$), the reference for the discovery of each quasar, and the mean SNR of the spectra in the three different wavelength ranges. The references and some notes are listed at the bottom of the table.

\section{Comparison with other quasar samples}\label{sec:qsosamples}
\begin{figure}
    \includegraphics[width=\linewidth]{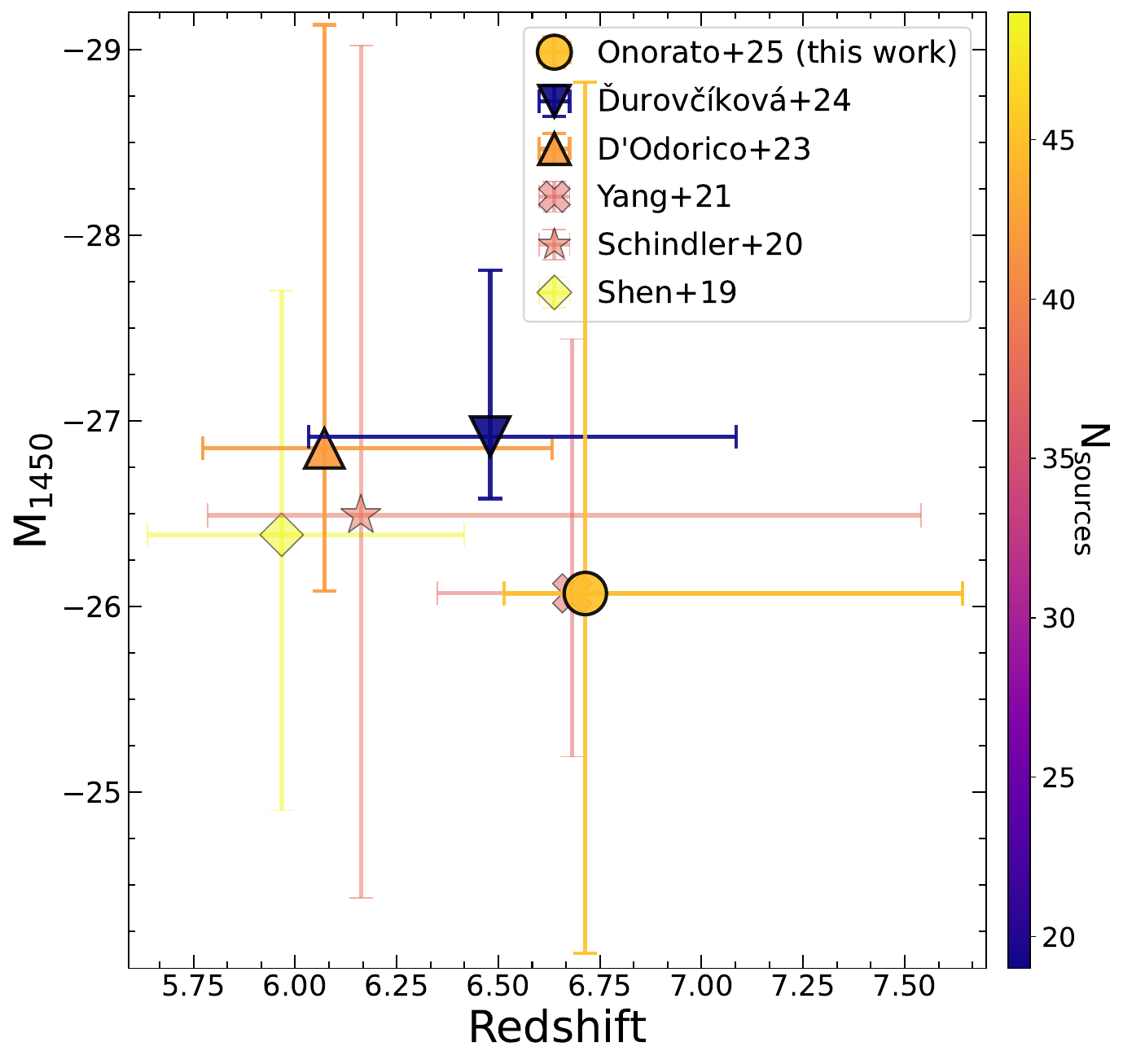}
    \caption{Comparison between the properties of the quasars in this work and those of the other high-$z$ samples \citep{Durovcikova2024, D'Odorico2023, Yang2021, Schindler2020, Shen2019}. We plot the median $z$ and $M_{1450}$ for all the catalogs, with markers color-coded with the number of sources included in each sample ($\rm{N_{sources}}$) as indicated by the colorbar on the right. The errorbars represent the extension of the samples in both $z$ and $M_{1450}$. The works where the spectra are publicly released (this work, \citealt{Durovcikova2024, D'Odorico2023}) are marked with thicker edges and brighter shades.}
    \label{fig:M1450_lit}
\end{figure}

It is important to discuss this work by comparing it to the present literature on other high-$z$ samples, such as \cite{Durovcikova2024, D'Odorico2023, Yang2021, Schindler2020, Shen2019}, where \cite{Durovcikova2024} and \cite{D'Odorico2023} are also spectroscopic data releases. The most relevant quantities used to characterize the samples are reported in Table \ref{tab:sample_compar}. In Figure \ref{fig:M1450_lit} we show the distribution of their median $z$ and $M_{1450}$ color-coded with the sample size and with the errorbars representing the extension of the samples in both $z$ and $M_{1450}$. The samples that also are public data releases are marked with thicker edges and brighter shades. We want to state clearly that for all the objects where the spectra are previously published, these are not the same reductions as in earlier publications, as all data are re-reduced in this work.


The sample that shows the most similarities with this work is the one in \cite{Yang2021}, with $31$ quasars in common in the redshift range $6.50<z<7.65$ out of the $37$ total sources. However, there are key differences between the two datasets. The major differences lie in the instruments used to collect the spectra: \cite{Yang2021} primarily used Gemini/GNIRS and Keck/NIRES, while in this work, a larger fraction of the data comes from VLT/X-Shooter, which offers better spectral resolution (see the $R$ values of each instrument reported in Table \ref{tab:instr_prop}) and, consequently, higher-quality spectra. Additionally, most of the spectra in \cite{Yang2021} have wavelength coverage only in the NIR arm, whereas our dataset includes broader coverage. These differences arise due to distinct scientific objectives. \cite{Yang2021} focused on central BH masses derived from the \ion{Mg}{II} emission lines, the Eddington ratio distribution, and chemical abundances in the rest-frame UV (such as the Fe II/\ion{Mg}{II} ratio). Consequently, their dataset did not prioritize coverage in the optical band. In contrast, we assembled a sample tailored for additional scientific goals, including the study of quasar proximity zones \citep{Onorato2025} and the Ly$\alpha$ forest, which require coverage in the VIS arm, a key feature provided by our dataset. Another important difference is that we have refined the redshift measurements for some quasars, incorporating more recent literature values and prioritizing [\ion{C}{II}] measurements when available. Additionally, unlike \cite{Yang2021}, which derived $M_{1450}$ from a best-fit power-law continuum, we independently compute these values from the spectra themselves by using the methodology described in Section \ref{subsec:M1450}.

The other three samples with only a few similarities with this work are: \cite{Durovcikova2024} with $8$ sources in common in the redshift range $6.53<z<7.09$ out of the $18$ total ones, \cite{Schindler2020} with $8$ quasars in common at $6.57<z<7.65$ out of the $38$ total ones, and \cite{D'Odorico2023} with only $6$ sources in common at $6.50<z<6.64$ out of the $42$ total ones. In the first sample mentioned, the spectra come almost entirely from Magellan/FIRE (\citealt{Simcoe2013}; $R = 6000$ if the slit is 0.6$^{\prime\prime}$ wide), which has a better data quality than the GNIRS and NIRES/MODS spectra in common, but worse than the X-Shooter ones. The other two samples have all the spectra acquired with VLT/X-Shooter.

Finally, there are no sources in common with \cite{Shen2019} as the redshift range ($5.71-6.42$) of their $50$ quasars does not overlap with the one in this work. All their spectra come from Gemini/GNIRS.

\begin{table*}
    \caption{Main properties of the sample in this paper, compared with those of other spectroscopic samples from the literature. The columns show respectively: the reference of the sample, the redshift range, the $M_{1450}$ range, the total number of quasars included in the sample (with the objects in common with this work), and the instruments used to take the spectra.}
    \label{tab:sample_compar}
    \begin{threeparttable}
    \begin{tabular}{lcccc}
        \hline
        Sample & $z$ range & $M_{1450}$ range & $\rm{N_{sources}}$ (in common) & Instruments\\
        \hline
        Onorato et al. (2025 - this work)$^{\rm{a}}$ & $6.50-7.65$ & $[-28.8,-24.1]$ & 45 & GNIRS/NIRES/X-Shooter/GMOS/LRIS/DEIMOS/MODS\\
        \cite{Durovcikova2024}$^{\rm{a}}$  & $6.03-7.08$ & $[-28.0,-26.5]$ & 18 (8) & FIRE/X-Shooter/MOSFIRE/ESI\\
        \cite{D'Odorico2023}$^{\rm{a}}$  & $5.77-6.63$ & $[-27.8,-25.8]$ & 42 (6) & X-Shooter\\
        \cite{Yang2021} & $6.30-7.65$ & $[-27.4,-25.2]$ & 37 (31) & GNIRS/NIRES/X-Shooter/FIRE/F2\\
        \cite{Schindler2020} & $5.78-7.54$ & $[-29.0,-24.4]$ & 38 (8) & X-Shooter\\
        \cite{Shen2019} & $5.71-6.42$ & $[-27.7,-24.9]$ & 50 & GNIRS\\
        \hline
    \end{tabular}
    \begin{tablenotes}
    \item[a] Public data release.
    \end{tablenotes}
    \end{threeparttable}
\end{table*}

\section{Composite Spectrum}\label{sec:composite}
\begin{figure*}
    \centering
    \includegraphics[width=\linewidth]{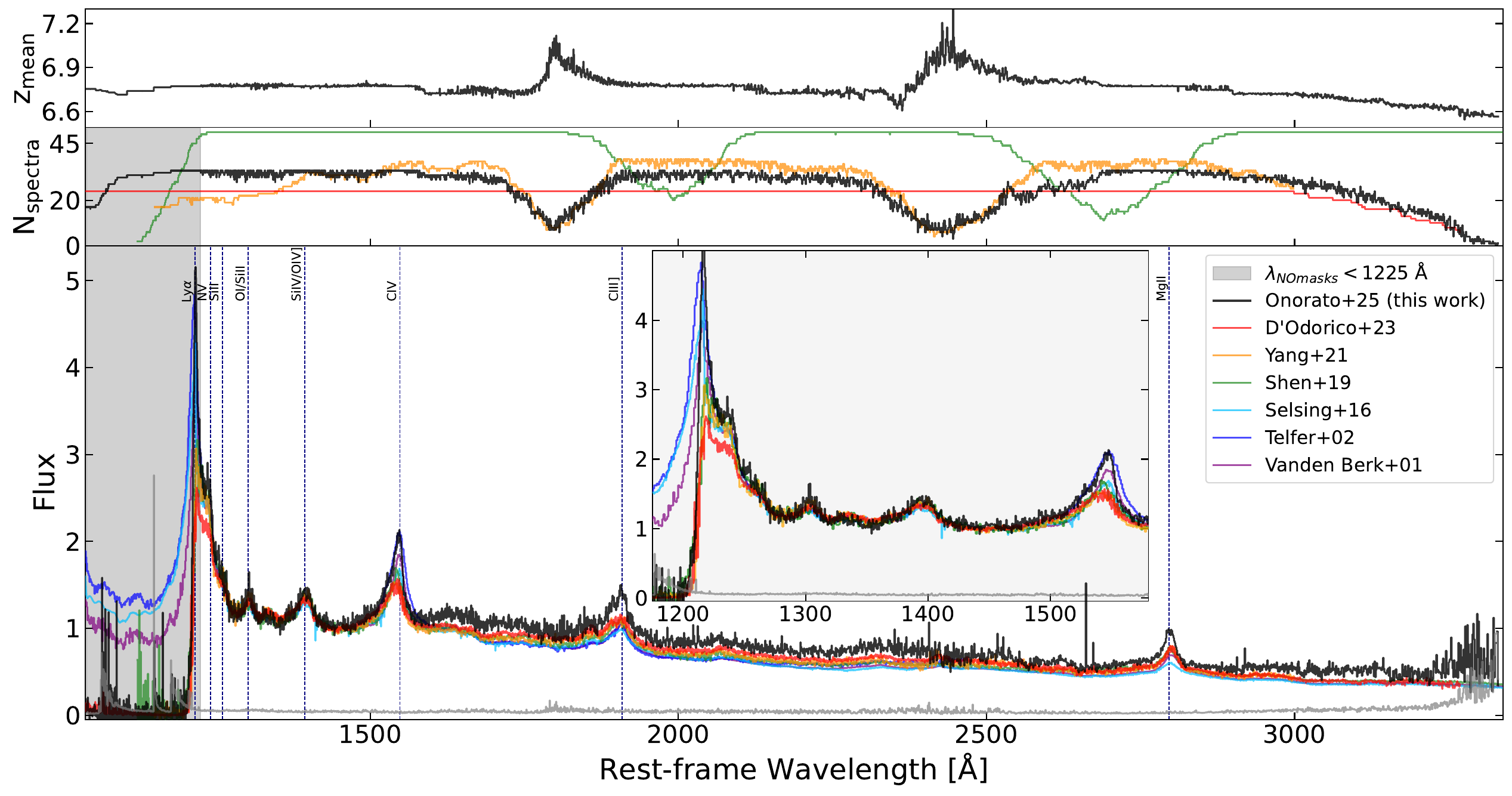}
        \caption{\textit{Bottom panel}: Composite spectrum of the $33$ non-BAL quasars in the sample (black) with its noise vector (grey), compared with several other composites from the literature (colored curves). All the composites are normalized to the continuum flux at 1450 {\angstrom}. The grey band for $\lambda<1225$ {\angstrom} highlights the mask-free region, where none of the masks described in the main text is applied. The main emission lines are shown as dashed blue lines. The inset panel shows a zoom-in of the region [1175, 1580] {\angstrom}, where the composites differ more. \textit{Middle panel}: Number of spectra that contribute to the composite realization at each wavelength for this work, \citealt{D'Odorico2023}, \citealt{Yang2019}, and \citealt{Shen2019} (same colors as in the bottom panel). \textit{Top panel}: Mean redshift of the quasars that contribute to the composite at each wavelength for this work.}
    \label{fig:composite}
\end{figure*}

In this section, we present a $z>6.5$ quasar composite spectrum based on this sample and compare it with other composite spectra known from the literature (see Figure \ref{fig:composite} and Table \ref{tab:composites}). We aim to study the average UV quasar spectral properties and their possible evolution through the different redshifts. We decide to exclude all the quasars that show BAL features in their spectra (flagged in Table \ref{tab:sample}) and then we generate the composite with $33$ out of the $45$ $z>6.5$ quasars in the sample. This choice comes from the fact that BAL features can distort the shape of the main emission lines, producing a prominent \ion{C}{IV} and many other high-ionization features, such as \ion{Si}{IV} and \ion{N}{V}.

We generate the composite spectrum following \cite{Selsing2016} as a guideline:
\begin{itemize}
    \item{We move the spectra to the rest-frame, dividing the wavelengths by $(1+z)$;}
    \item{We define a new wavelength grid: [1040, 3332] {\angstrom} in steps of $dv_{\rm{pix}}$=110 km s$^{-1}$ (set to be coarser than the coarsest pixel-scale, in velocity, among all the instruments used to create the sample);}
    \item{We re-bin the spectra into the new wavelength grid;}
    \item{We normalize the spectra to the continuum flux at 1450 {\angstrom} rest-frame, where there are no strong broad lines or iron emission;}
    \item{Only for $\lambda>1225$ {\angstrom}, we apply the following masks to improve the overall quality of the composite, without affecting its natural shape in the Ly$\alpha$ region:
        \begin{enumerate}
        \item Telluric transmission $>0.5$: we use the individual \texttt{PypeIt} telluric model fits to mask out the telluric regions;
        \item SNR $>0.5$: to mask the flux where the SNR is very low (and hence the noise is high);
        \item $\sigma$ [$10^{-17}$ erg s$^{-1}$cm$^{-2}${\angstrom}$^{-1}$] $<1.5$: to mask the flux where the noise is high (i.e., in those regions close to the right edge of the wavelength coverage of the spectra, where the noise increases exponentially);
        \item Flux [$10^{-17}$ erg s$^{-1}$cm$^{-2}${\angstrom}$^{-1}$] $<40$: to mask out the outliers (e.g., hot pixels or sky lines not well subtracted during the spectra reduction).
        \end{enumerate}}
    \item{We create the composite spectrum as a weighted mean of the individual spectra. The weight assigned to each spectrum at each wavelength is given by Equation \ref{eq:weights}. The different factors in the equation are: the good pixel mask (gpm; see Table \ref{tab:spectra}) that removes bad pixels from the calculation, the combination of the four masks defined above to exclude low SNR regions, high-noise areas, and spurious outliers (masks$_{\rm{tot}}$), and the number of spectra contributing at each wavelength (n$_{\rm{used}}$). Thus, the composite spectrum at every wavelength is the sum of the fluxes of the individual spectra at that wavelength, each multiplied by its weight, and divided by the sum of the weights.
    \begin{equation}
    \label{eq:weights}
        \rm{weights} = \frac{\rm{gpm} \cdot \rm{masks_{tot}}}{n_{\rm{used}}}
    \end{equation}}
\end{itemize}

\begin{table}
    \caption{Composite spectrum of the $33$ non-BAL quasars in the sample. Wavelengths are in the rest-frame and units of {\angstrom}. Flux density units are arbitrary, normalized to the rest-frame 1450 {\angstrom} continuum flux. The third column is the noise vector. The fourth column indicates the number of quasar's spectra contributing to the composite at each wavelength. The last column is the mean redshift that contributes to the composite at each wavelength. The entire table data is available online.}
    \centering
    \begin{tabular}{lcccc}
    \hline
      Wavelengths [{\angstrom}] & Flux [$\rm{F_{\lambda}}$] & Error & N spec & Mean $z$\\
    \hline
      1040.00000 & 0.07235 & 0.03021 & 18 & 6.73951 \\
      1040.38167 & 0.14979 & 0.02575 & 18 & 6.73951 \\
      1040.76347 & 0.24631 & 0.02212 & 18 & 6.73951 \\
      ... & & ... & & ... \\
      1600.55798 & 1.36016 & 0.04412 & 30 & 6.72212 \\
      ... & & ... & & ... \\
      2096.78621 & 0.97827 & 0.04707 & 33 & 6.76699 \\
      ... & & ... & & ... \\
      3250.71769 & 0.78563 & 0.14352 & 13 & 6.62963 \\
      ... & & ... & & ... \\
      \hline
    \end{tabular}
    \label{tab:composite_det}
\end{table}

The composite spectrum that we obtain is available online as explained in the Data Availability Section, and an example of the format of the \texttt{FITS} file is shown in Table \ref{tab:composite_det}. 
For $\lambda \in$ [1700, 1900] {\angstrom} and $\lambda \in$ [2300, 2600] {\angstrom} (rest-frame), the quality of the composite decreases because of the presence of strong telluric absorption at $\lambda \in$ [13500, 14150] {\angstrom} and $\lambda \in$ [18200, 19300] {\angstrom} (observed frame). Indeed, at the wavelengths of the highest absorption, only $\simeq 5-10$ spectra contribute to the composite because of the narrow redshift range of our sample. The quality also decreases for $\lambda > 3100$ {\angstrom}, where the number of contributing spectra starts to get lower.

In the bottom panel of Figure \ref{fig:composite}, we compare our composite with those from \cite{D'Odorico2023, Yang2021, Shen2019, Selsing2016, Telfer2002, VandenBerk2001}; for comparison purposes, all the composites are normalized at 1450 {\angstrom}, showing a better match among their continuum shape for $\lambda<1580$ {\angstrom} rather than at longer wavelengths, where there is a visible offset between the various continua. For a better comparison with our work, we consider the non-BAL composite from \cite{D'Odorico2023}.
We also report the main emission lines in this wavelength range, and an inset panel with a zoom-in of the region [1175, 1580] {\angstrom}, where we can see differences in the Ly$\alpha$ and \ion{C}{IV} lines, despite the good overlap among their continuum shape. In the middle panel of Figure \ref{fig:composite}, we display the number of spectra that are contributing to the composite at each wavelength for our work, \cite{D'Odorico2023}, \cite{Yang2021}, and \cite{Shen2019} (the only three for which this piece of information is available); and finally, in the top panel, we show the distribution of the mean quasar redshift that contributes to the composite at each wavelength. We summarize the properties of our composite and the comparison ones in Table \ref{tab:composites}.

\begin{table*}
    \caption{Main properties of the non-BAL composite spectrum created from the sample in this paper, compared with those of other composites. The columns show respectively: the reference of the composite, the redshift range of the sample, the $M_{1450}$ range for the sample, the total number of quasars included in the sample, and the instruments used to take the spectra.}
    \label{tab:composites}
    \begin{threeparttable}
    \begin{tabular}{lclcc}
        \hline
        Composite & $z$ range & $M_{1450}$ range & $\rm{N_{tot}}$ quasars & Instruments\\
        \hline
        Onorato et al. (2025 - this work) & $6.50-7.55$ & $[-27.4,-25.2]$ & 33 & GNIRS/NIRES/X-Shooter/GMOS/LRIS/MODS\\
        \cite{D'Odorico2023} & $5.77-6.63$ & $[-27.8,-25.8]$ & 24$^{\rm{a}}$ & X-Shooter\\
        \cite{Yang2021} & $6.50-7.65$ & $[-27.4,-25.2]$ & 38$^{\rm{b}}$ & GNIRS/NIRES/X-Shooter/FIRE/F2\\
         \cite{Shen2019} & $5.71-6.42$ & $[-27.7,-24.9]$ & 50 & GNIRS\\
         \cite{Selsing2016} & $1.12-2.10$ & $[-28.1,-27.2]^{\rm{c}}$ & 7 & X-Shooter\\
         \cite{Telfer2002} & $0.33-3.60$ & $[-27.7,-24.7]^{\rm{c}}$ & 184 & FOS/GHRS/STIS\\
         \cite{VandenBerk2001} & $0.044-4.789$ & $[-25.8,-22.8]^{\rm{d}}$ & 2204 & SDSS\\
        \hline
    \end{tabular}
    \begin{tablenotes}
    \item[a] Excluding the BALs.
    \item[b] Where 31 come from \cite{Yang2021}, and 7 from \cite{Schindler2020}.
    \item[c] From \cite{Lusso2015}, $M_{1450} = M_{i} (z = 2) + 1.28$.
    \item[d] Converted from $17.5<r'<20.5$, with $z_{\rm{median}}=1.253$.
    \end{tablenotes}
    \end{threeparttable}
\end{table*}

The discrepancies highlighted in the inset panel of Figure \ref{fig:composite} are most likely due to differences in luminosity among the samples (see the $M_{1450}$ column in Table \ref{tab:composites}), as we know that many emission line properties are functions of quasar luminosity. This is the case of \ion{C}{IV}, a high-ionization line visible in the composites, where a decreasing equivalent line width (EW) is expected with increasing luminosity, according to the so-called Baldwin effect \citep{Baldwin1977}. The biggest differences are noticeable in the strength of the \ion{C}{IV} emission line when comparing our composite with the one from the most bright samples, such as \cite{Selsing2016}, \cite{Shen2019} and \cite{D'Odorico2023}.
This effect is also confirmed by the test described in Section \ref{sec:composite_bins_mag}, where we divide the sample of non-BAL quasars into two luminosity bins and create a separate composite from each one. In Figure \ref{fig:composite_comp_magbins}, we can notice the different EW of the Ly$\alpha$, \ion{C}{IV}, and \ion{Mg}{II} emission lines due to the difference in $M_{1450}$ of the two sub-samples.
The different strength of the Ly$\alpha$ line between our sample (or high-$z$ in general) and low-$z$ ones is due to the increasing absorption from the IGM toward higher redshift. The overall continuum slope of the composite is consistent with those from the literature, showing that the same spectral features are preserved in quasars at different $z$ ranges.

In the next sections, we discuss an alternative version of the composite obtained including the BAL quasars, and two tests performed by dividing the $33$ non-BAL quasars into two redshift bins first, and into two $M_{1450}$ bins later.

\subsection{Including BAL quasars}\label{sec:composite_bal}
\begin{figure*}
    \centering
    \includegraphics[width=\linewidth]{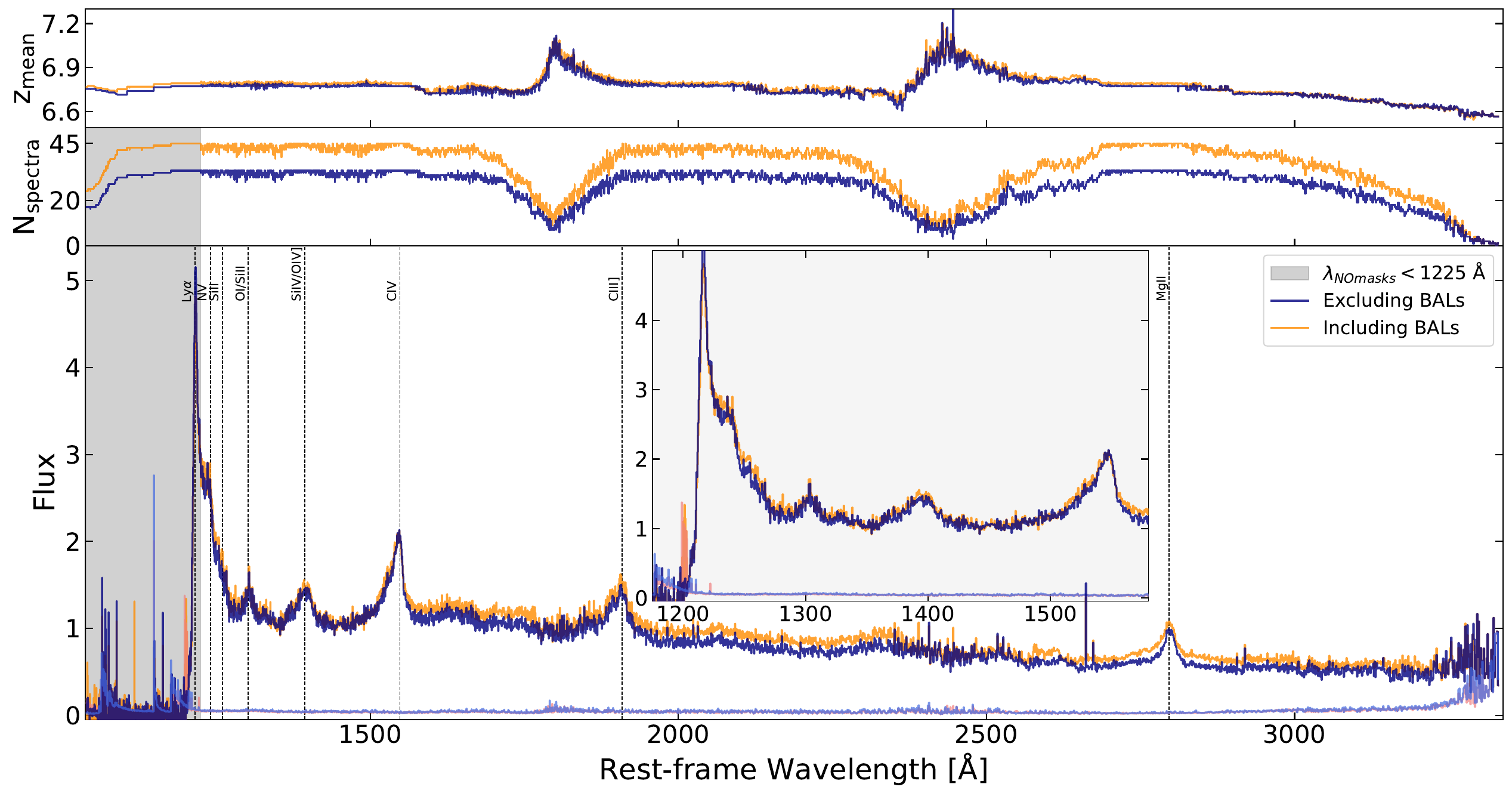}
    \caption{\textit{Bottom panel}: Composite spectrum of the $33$ non-BAL quasars in the sample (dark blue curve, which is the same as the black curve in Figure \ref{fig:composite}), compared with the composite spectrum obtained including the $12$ BAL quasars (orange curve) which are flagged in Table \ref{tab:sample}, and also listed in Section \ref{sec:sample}. There are no significant differences between the two versions of the composite. The rest of the plot shows the same quantities already described in Figure \ref{fig:composite}.}
    \label{fig:composite_comp}
\end{figure*}

As already stated at the beginning of Section \ref{sec:composite}, BAL features can affect the shape of a quasar spectrum. For this reason, we decide to exclude the contribution of all the BAL quasars in the sample to the creation of the final composite spectrum. However, for completeness purposes, we also create another version of the composite, this time considering all the quasars in the sample, so including the BALs. We show the comparison between the two versions in Figure \ref{fig:composite_comp}, where the blue curve is the composite created from the $33$ non-BAL quasars in the sample, and the orange curve is the one obtained including also the $12$ BALs flagged in Table \ref{tab:sample}, and listed in Section \ref{sec:sample}. The rest of the plot shows the same quantities already described in Figure \ref{fig:composite}.
From this comparison, we can see that there are no strong differences between the two versions of the composite. Both \cite{D'Odorico2023} and \cite{Yang2021} make the same investigation on the BAL contribution to their composite spectrum. They agree with this study that BALs do not cause significant differences but, unlike in this work, decide to include them in their final composite anyway.

\subsection{Dividing the sample into two redshift bins}\label{sec:composite_bins}
\begin{figure*}
    \centering
    \includegraphics[width=\linewidth]{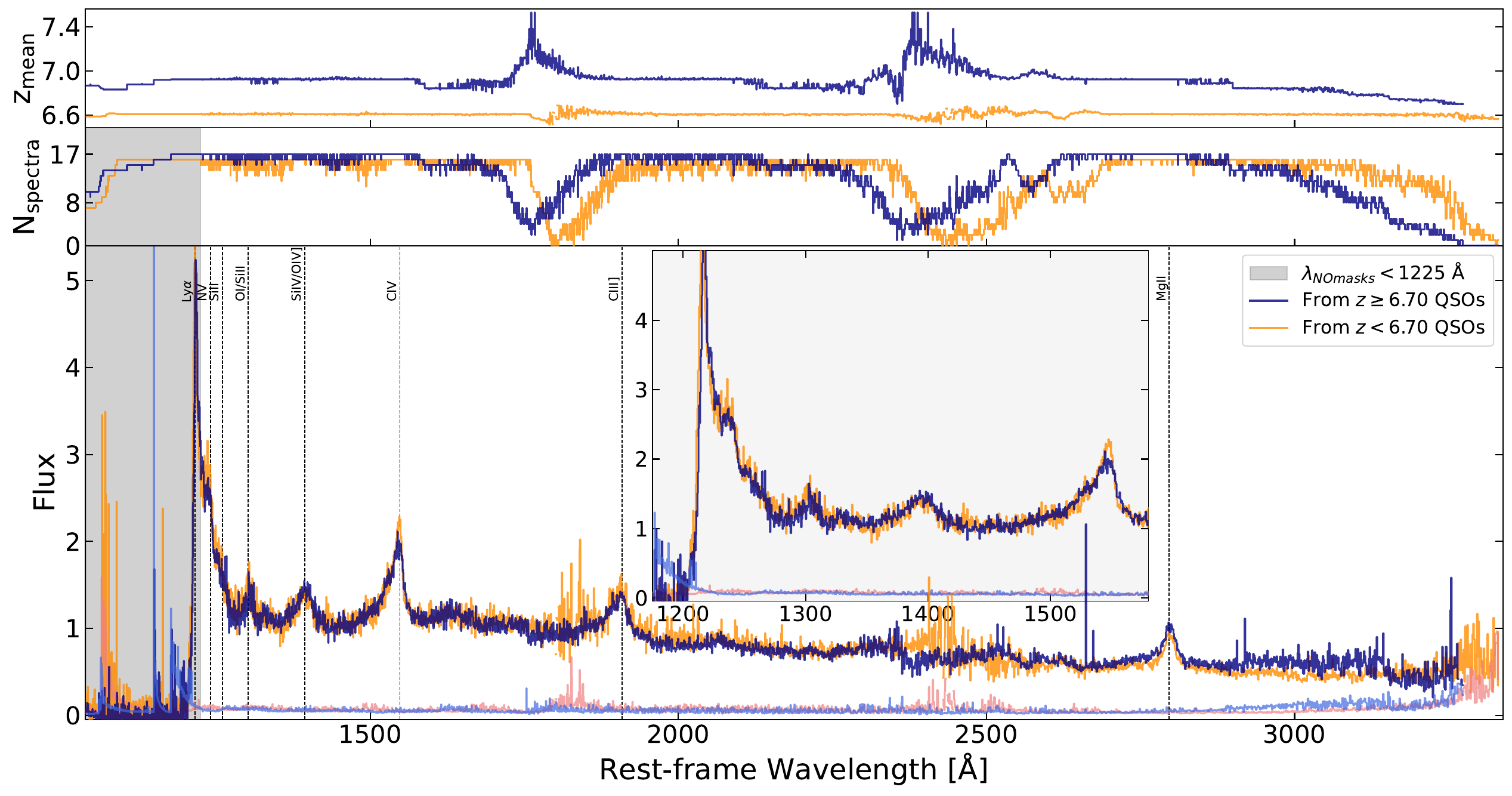}
    \caption{\textit{Bottom panel}: Composite spectrum of the $17$ spectra having $z \geq 6.70$ (dark blue curve), compared with the composite spectrum obtained from the $16$ spectra with $z < 6.70$ (orange curve); where $z_{\rm{median,non-BAL}}=6.70$ is the median redshift of the $33$ non-BAL quasars. The rest of the plot shows the same quantities already described in Figure \ref{fig:composite}.}
    \label{fig:composite_comp_zbins}
\end{figure*}

To check whether we are averaging any features in the final composite spectrum because of the relatively wide redshift distribution of the quasars in the sample, we perform another test dividing the $33$ non-BAL quasars into two redshift bins and creating a composite spectrum from the spectra in each bin. If any "$z$-related" features are present, they might arise when dividing the sample into two groups based on the redshift. To account for enough statistics, we consider the median redshift of the $33$ non-BALs ($z_{\rm{median,non-BAL}}=6.70$) and create a version of the composite from the $17$ spectra having $z \geq 6.70$ and another version from the $16$ spectra with $z < 6.70$. The two versions are shown in Figure \ref{fig:composite_comp_zbins}, where the blue curve is the composite created from the quasars at $z \geq 6.70$, and the orange curve is the one obtained from the quasars at $z < 6.70$. The rest of the plot shows the same quantities already described in Figure \ref{fig:composite}.

From this comparison, we notice that the high-$z$ composite has a shallower slope and thus a redder continuum. A possible interpretation of this would be a larger number of mini BAL absorption at higher redshift. \cite{Bischetti2022, Bischetti2023} found that the BAL fraction in $z \geq 6$ quasars is $2-3$ times higher than in quasars at $z \simeq 2-4.5$. The presence of BALs correlates with a redder continuum, probably due to dust attenuation. If this idea is correct, it would explain why the Ly$\alpha$ and \ion{C}{IV} lines look weaker.
Alternatively, another explanation to justify the weaker Ly$\alpha$ could be the presence of more neutral hydrogen in the IGM causing more absorption and suggesting the presence of statistical IGM damping wings \citep{Durovcikova2024}.

\subsection{Dividing the sample into two magnitude bins}\label{sec:composite_bins_mag}
\begin{figure*}
    \centering
    \includegraphics[width=\linewidth]{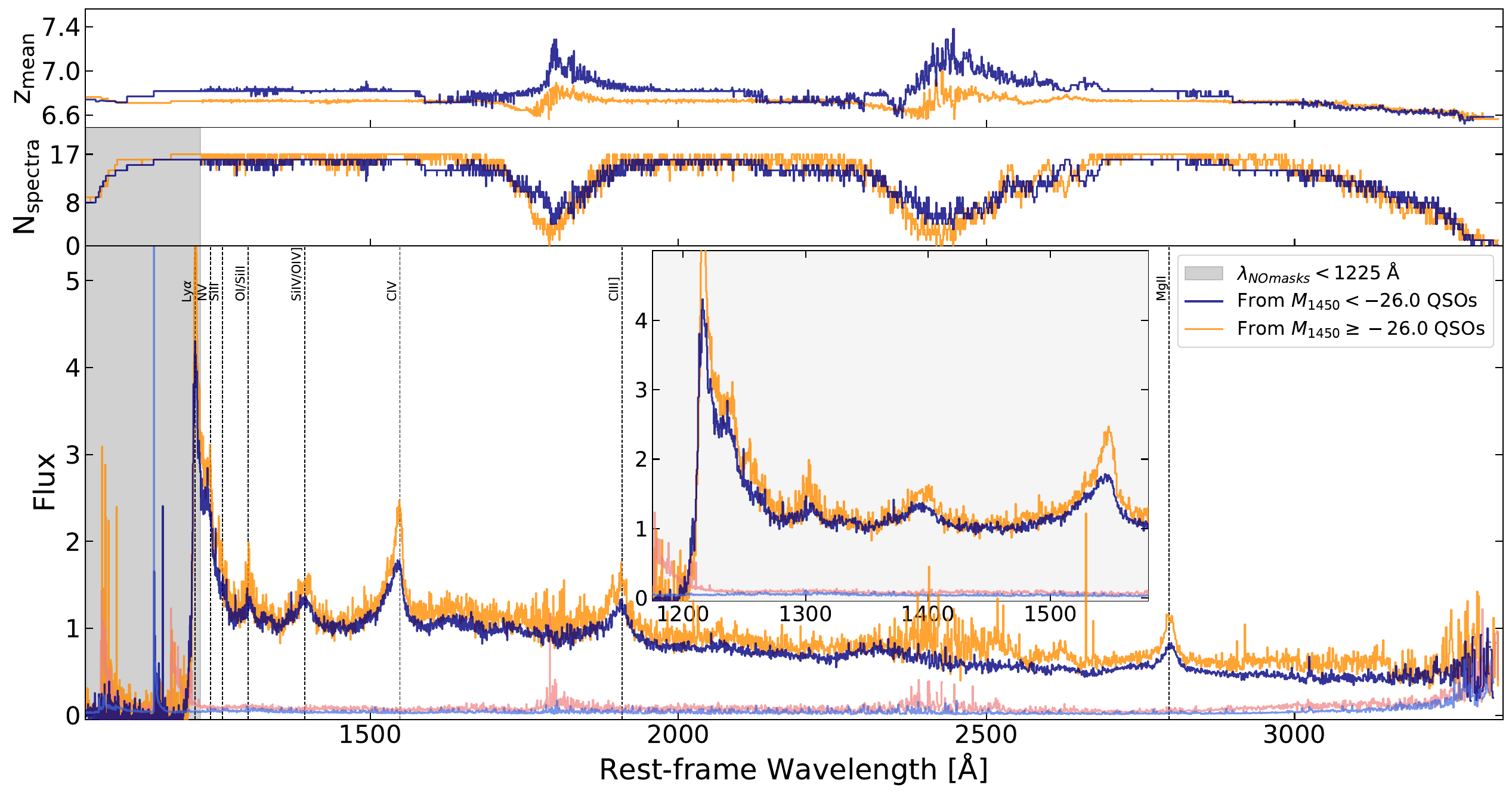}
    \caption{\textit{Bottom panel}: Composite spectrum of the $16$ spectra having $M_{1450} < -26.0$ (dark blue curve), compared with the composite spectrum obtained from the $17$ spectra with $M_{1450} \geq -26.0$ (orange curve); where $M_{1450,\rm{median,non-BAL}}=-26.0$ is the median $M_{1450}$ of the $33$ non-BAL quasars. The rest of the plot shows the same quantities already described in Figure \ref{fig:composite}.}
    \label{fig:composite_comp_magbins}
\end{figure*}

The last test we perform on the composite consists of dividing the $33$ non-BAL quasars into two $M_{1450}$ bins and creating a composite spectrum from the spectra in each bin. As before, to account for enough statistics, we consider the median $M_{1450}$ of the $33$ non-BALs ($M_{1450,\rm{median,non-BAL}}=-26.0$) and create a version of the composite from the $16$ spectra having $M_{1450} < -26.0$ and another version from the $17$ spectra with $M_{1450} \geq -26.0$. The two versions are shown in Figure \ref{fig:composite_comp_magbins}, where the blue curve is the composite created from the quasars having $M_{1450} < -26.0$, and the orange curve is the one obtained from the quasars having $M_{1450} \geq -26.0$. The rest of the plot shows the same quantities already described in Figure \ref{fig:composite}.

The differences in the strength of the emission lines come from the Baldwin effect \citep{Baldwin1977}, already discussed in Section \ref{sec:composite}, proving that quasars' luminosity anti-correlates with their emission lines strength. From this test, we show how this effect is still visible at early times.

\section{Summary}\label{sec:summary}
In this section, we discuss the conclusions of this paper, the quantitative analysis performed on the sample, and future work. We summarize the two main results below.

\begin{itemize}
    \item We report and release a sample of $45$ quasars at $6.50 < z \leq 7.64$ ($z_{\rm{median}}=6.71$), in the magnitude range $-28.82 \leq M_{1450} \leq -24.13$ ($M_{1450,\rm{median}} \simeq -26.1$), where $12$ out of the $45$ quasars show BAL features in their spectrum, and $3$ are new unpublished quasars \citep{Banados2025, Belladitta2025}. This represents the largest medium/moderate-resolution released sample of quasars at high redshift from ground-based instruments. The optical and NIR spectroscopy were obtained using the Gemini/GNIRS, Keck/NIRES, VLT/X-Shooter, Gemini/GMOS, Keck/DEIMOS, Keck/LRIS, and LBT/MODS instruments. The data in this sample allow us to estimate the $M_{1450}$ of the quasars directly from the spectra, and to determine their quality from the SNR:

    \begin{enumerate}
        \item After scaling the flux and inverse variance of each spectrum considering the J, Y, or $\rm{K_{p}}$ band magnitude of every quasar, we calculate $M_{1450}$ from these new flux-scaled spectra (see Section \ref{subsec:M1450} and Appendix \ref{app:bal}). These values may be relevant in the course of the analysis on quasar proximity zones \citep{Onorato2025}, in case we have to correct the measurements for the luminosity of the quasar.
        \item We compute the SNR of the spectra in the sample, considering three different wavelength ranges which exclude the telluric regions: $[11000, 13400]$, $[14500, 17950]$, and $[19650, 22400]$ {\angstrom} (see Section \ref{sec:snr} and Appendix \ref{app:snr}). We want to test the quality of the spectra and give an idea of how much a certain spectrum can be trusted.
    \end{enumerate}
    
    \item We create a $z > 6.5$ quasar composite spectrum using $33$ out of the $45$ quasars of this sample: we exclude the BAL quasars to avoid biases in the analysis due to absorption features in their spectra. We compare the composite with others from low and high redshift quasars samples from the literature. No significant redshift evolution is found for either broad UV emission lines or quasar continuum slopes (see Section \ref{sec:composite}). An alternative version of the composite, created considering also the BAL quasars, is described in Section \ref{sec:composite_bal}. We notice that there are no strong differences between the two versions. Other two tests are discussed in Section \ref{sec:composite_bins} and \ref{sec:composite_bins_mag}. In the first one, we divide the sample of $33$ non-BAL quasars into two redshift bins (17 quasars at $z \geq 6.70$ and $16$ at $z<6.70$) and create a composite from the spectra in each bin. There are small differences in the continuum shape and the strength of Ly$\alpha$ and \ion{C}{IV} emission lines, visible in Figure \ref{fig:composite_comp_zbins}, that we tentatively interpret as due to the presence of subtle BAL absorption at higher $z$ \citep{Bischetti2022, Bischetti2023}. Alternatively, statistical damping wings \citep{Durovcikova2024} could also justify the weaker Ly$\alpha$ observed at higher $z$, while the difference in \ion{Mg}{II} could arise from changes in the average black hole mass or accretion rates. In the second test, we divide the sample of $33$ non-BAL quasars into two $M_{1450}$ bins (17 quasars at $M_{1450} \geq -26.0$ and $16$ at $M_{1450}<-26.0$) and create a composite from the spectra in each bin. In Figure \ref{fig:composite_comp_magbins}, we see how weaker emission lines are associated with brighter quasars, confirming the presence of the Baldwin effect \citep{Baldwin1977}, which still holds at early times.
\end{itemize}

Starting from the sample presented in this paper, in subsequent works we will reconstruct the quasar's intrinsic blue side from the observed red side, with PCA continuum modelling (e.g., \citealt{Davies2018a, Davies2018b}; \citealt{Bosman2021}), to study their proximity zones \citep{Onorato2025}, and Ly$\alpha$ damping wings (\citealt{Miralda-Escude1998}; \citealt{Davies2018a}; \citealt{Durovcikova2024}; \citealt{Greig2024}). In particular, \cite{Hennawi2024} introduce a new inference approach for analyzing the IGM damping wings, deriving a single Bayesian likelihood for the entire spectrum; while \cite{Kist2024}, quantify the precision with which these IGM damping wings analyzed with the new method can measure astrophysical parameters such as $\langle x_{\rm{HI}} \rangle$ and $t_{\rm{Q}}$, and the dependence of this precision on the properties of the spectra analyzed.
Our final goal will be to impose more stringent constraints on $\langle x_{\rm{HI}} \rangle$ during the EoR and the radiative efficiency of the earliest SMBHs.

\section*{Acknowledgements}
This work is based in part on observations obtained at the international Gemini Observatory, a program of NSF NOIRLab, which is managed by the Association of Universities for Research in Astronomy (AURA) under a cooperative agreement with the U.S. National Science Foundation on behalf of the Gemini Observatory partnership: the U.S. National Science Foundation (United States), National Research Council (Canada), Agencia Nacional de Investigaci\'{o}n y Desarrollo (Chile), Ministerio de Ciencia, Tecnolog\'{i}a e Innovaci\'{o}n (Argentina), Minist\'{e}rio da Ci\^{e}ncia, Tecnologia, Inova\c{c}\~{o}es e Comunica\c{c}\~{o}es (Brazil), and Korea Astronomy and Space Science Institute (Republic of Korea).

Some of the data presented herein were obtained at Keck Observatory, which is a private 501(c)3 non-profit organization operated as a scientific partnership among the California Institute of Technology, the University of California, and the National Aeronautics and Space Administration. The Observatory was made possible by the generous financial support of the W. M. Keck Foundation.
The authors wish to recognize and acknowledge the very significant cultural role and reverence that the summit of Maunakea has always had within the Native Hawaiian community. We are most fortunate to have the opportunity to conduct observations from this mountain.
This research has made use of the Keck Observatory Archive (KOA), which is operated by the W. M. Keck Observatory and the NASA Exoplanet Science Institute (NExScI), under contract with the National Aeronautics and Space Administration.

This work is based in part on observations made with ESO telescopes at the La Silla Paranal Observatory.

This paper also uses data based on observations obtained at the LBT, an international collaboration among institutions in the United States, Italy and Germany. The LBT Corporation partners are: The University of Arizona on behalf of the Arizona university system; Istituto Nazionale di Astrofisica, Italy;  LBT Beteiligungsgesellschaft, Germany, representing the Max Planck Society, the Astrophysical Institute Potsdam, and Heidelberg University; The Ohio State University; The Research Corporation, on behalf of The University of Notre Dame, University of Minnesota and University of Virginia.

%
%
We acknowledge the use of the \texttt{PypeIt} data reduction package.

We acknowledge helpful conversations with the ENIGMA group at UC Santa Barbara and Leiden University.
SO is grateful to Ben Wang and Suk Sien Tie for help with some of the data reductions, and to Elia Pizzati and Caitlin Doughty for comments on an early version of the manuscript.

JFH acknowledges support from the European Research Council (ERC) under the European Union’s Horizon 2020 research and innovation program (grant agreement No 885301), from the National Science Foundation (NSF) under Grant No. 2307180, and from NASA under the Astrophysics Data Analysis Program (ADAP, Grant No. 80NSSC21K1568).
JTS is supported by the Deutsche Forschungsgemeinschaft (DFG, German Research Foundation) - Project number 518006966.
F.Wang acknowledges support from NSF award AST-2513040.
CM acknowledges support from Fondecyt Iniciacion grant 11240336 and the ANID BASAL project FB210003.
E.P.F. is supported by the international Gemini Observatory, a program of NSF’s NOIRLab, which is managed by the Association of Universities for Research in Astronomy (AURA) under a cooperative agreement with the National Science Foundation, on behalf of the Gemini partnership of Argentina, Brazil, Canada, Chile, the Republic of Korea, and the United States of America.

\section*{Data Availability}
\subsection*{Data archives}
The raw Gemini data (both GNIRS and GMOS) can be searched and downloaded from the Gemini Observatory Archive at this link: \url{https://archive.gemini.edu/searchform}. The user needs to set the Instrument used and the coordinates of the target (RA and Dec). Additionally, we notice that sometimes there are superimposed features like vertical striping, horizontal banding, and quadrant offsets on GNIRS data. We use the \texttt{CLEANIR}\footnote{\url{https://www.gemini.edu/instrumentation/niri/data-reduction}} Python routine to remove these artifacts.

The raw Keck data (all NIRES, DEIMOS, and LRIS) can be downloaded from a Basic Search on the Keck Observatory Archive (KOA) at this link: \url{https://koa.ipac.caltech.edu/cgi-bin/KOA/nph-KOAlogin}. The user needs to select the Instrument used, set the Object Name or Location, and decide how to return the results. In case we want to download multiple targets, we can look for More Search Options, select the desired Instruments, and add a file in the Multiple Object Table File section. A code to create such a table is provided in the GitHub repository
storing the relevant tables and codes used in this paper, whose link is provided in Section \ref{sec:dataredux}.

The raw ESO data (VLT/X-Shooter) can be downloaded from the ESO Observatory Archive at this link: \url{https://archive.eso.org/eso/eso_archive_main.html}. The user can type the coordinates of the target (RA and Dec), set the Instrument used, and select the category of data desired. In case we want to download multiple targets, we can add a file in the List of Targets section. A code to create such a file is provided in the GitHub repository already mentioned.

The raw LBT data (MODS) come from private projects present in the LBT Archive at this link: \url{http://archive.lbto.org/}. The user needs to set the coordinates of the target (RA and Dec), the Instrument used, and to authenticate for the project.

More (and fewer) details can always be inserted in the archives when searching for targets.

\subsection*{Data reduction and analysis}
The files to reduce the spectroscopic data with \texttt{PypeIt} (\texttt{.pypeit}, \texttt{.flux}, \texttt{.coadd1d}, and \texttt{.tell}), the \texttt{.csv} tables (containing information such as the redshift of the quasars, photometric details or SNR), and the relevant codes to reproduce the results in this paper are all stored in the GitHub repository previously mentioned.

\subsection*{Final FITS files (spectra and composites)}
The data underlying this article are available in the article and in its online supplementary material.



\bibliographystyle{mnras}
\bibliography{ref} 




\appendix

\section{Flux scaling with J, Y, or $\rm{K_{p}}$ band photometry}\label{app:fluxscale}
As already mentioned in Section \ref{subsec:fluxscale}, the reduced spectrum of every quasar in this sample is scaled using the J, Y, or $\rm{K_{p}}$ band photometry. This is because the flux calibration of the spectra performed with \texttt{PypeIt} is relative and, while we are still looking for a definitive absolute flux calibration method, we need to tie the spectroscopic measurements with the photometric ones, which are more accurate in terms of fluxes.

\begin{figure*}
    \includegraphics[width=0.66\linewidth]{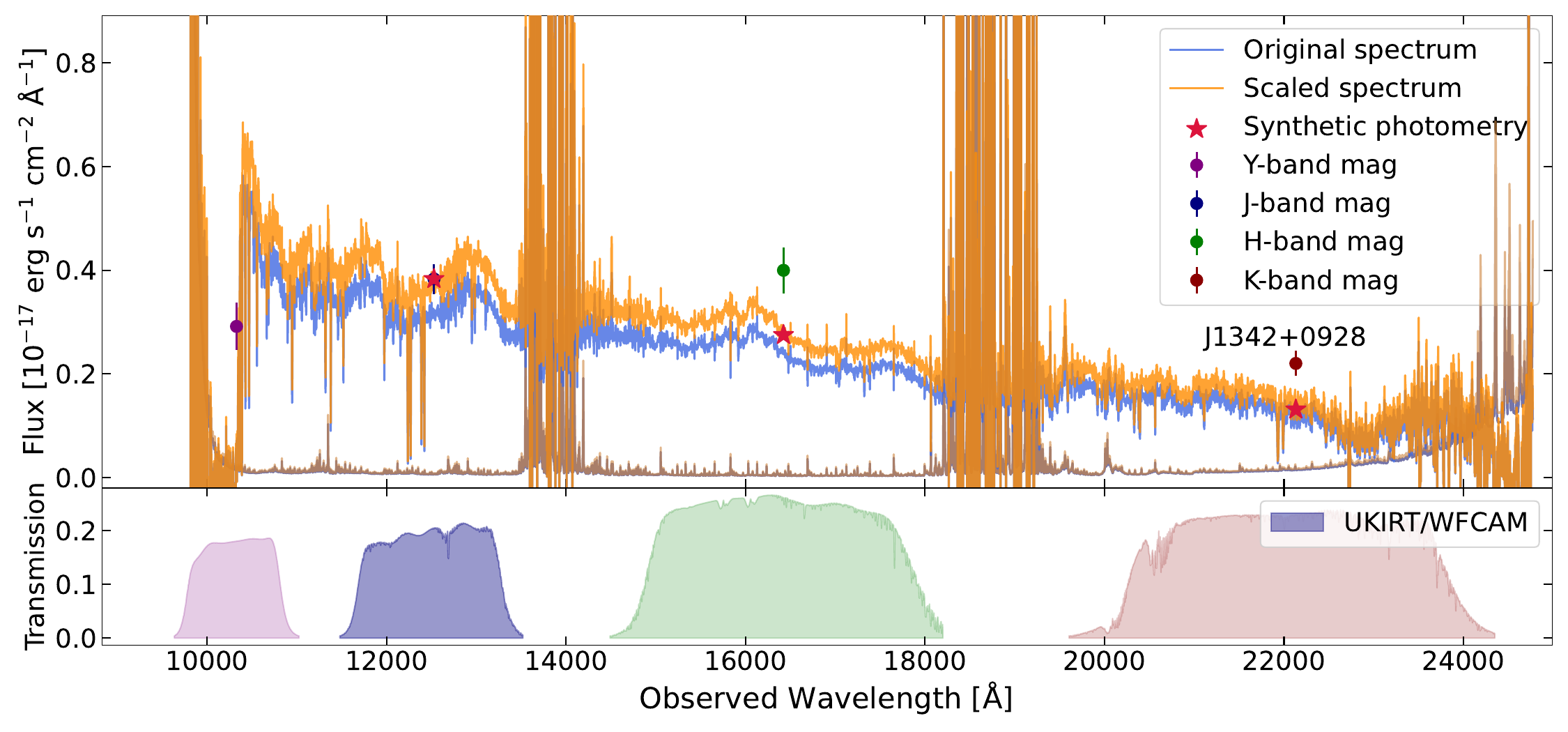}
    \includegraphics[width=0.66\linewidth]{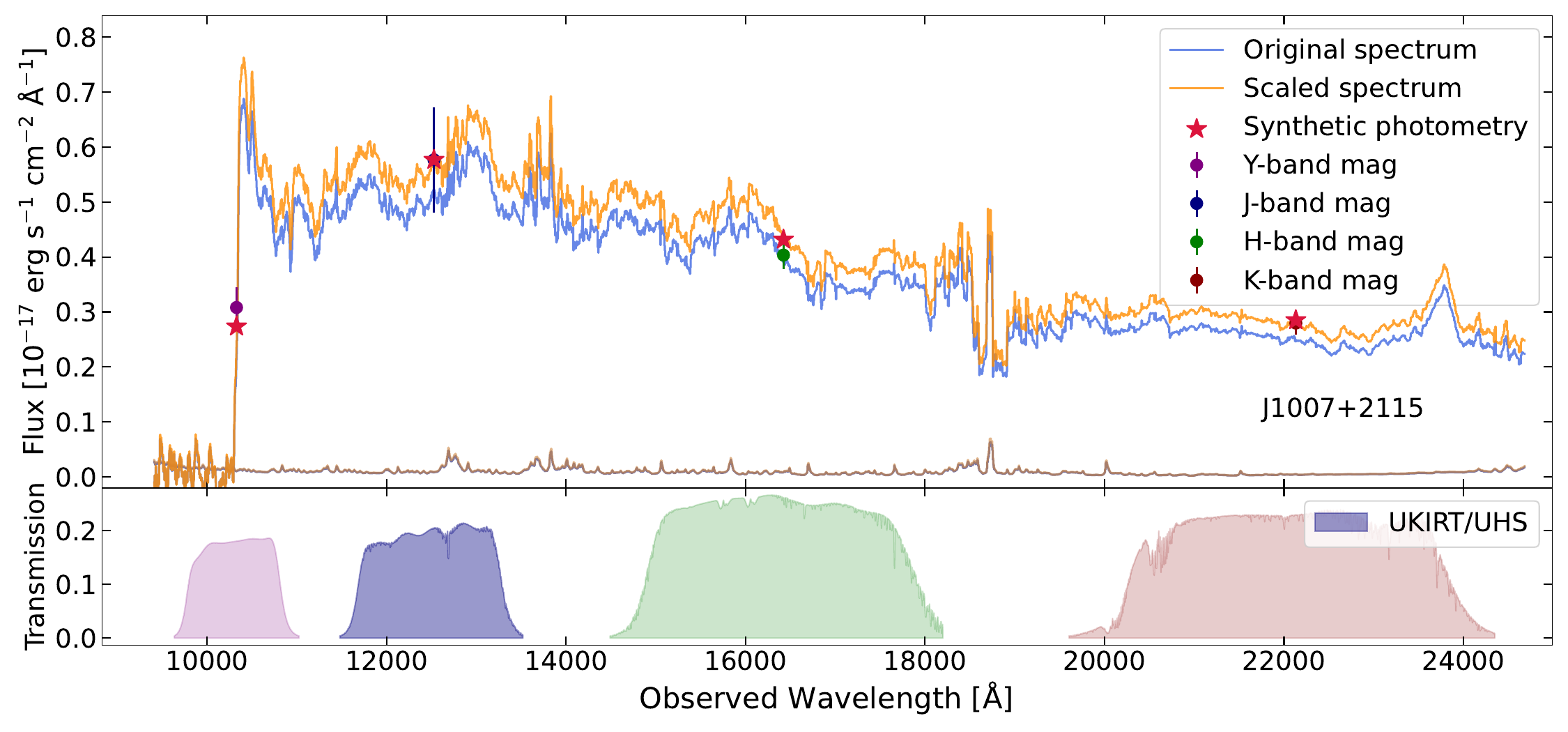}
    \includegraphics[width=0.66\linewidth]{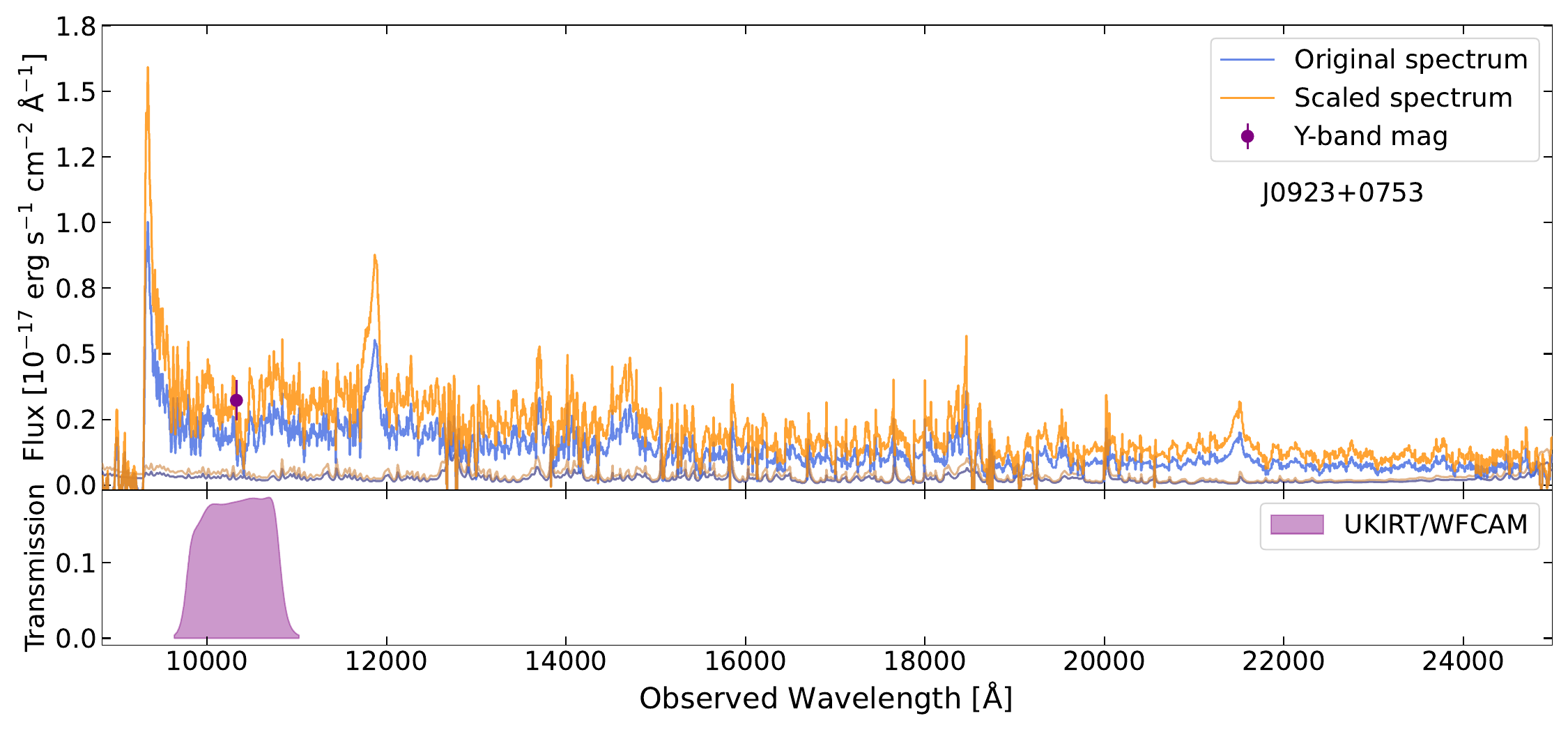}
    \includegraphics[width=0.66\linewidth]{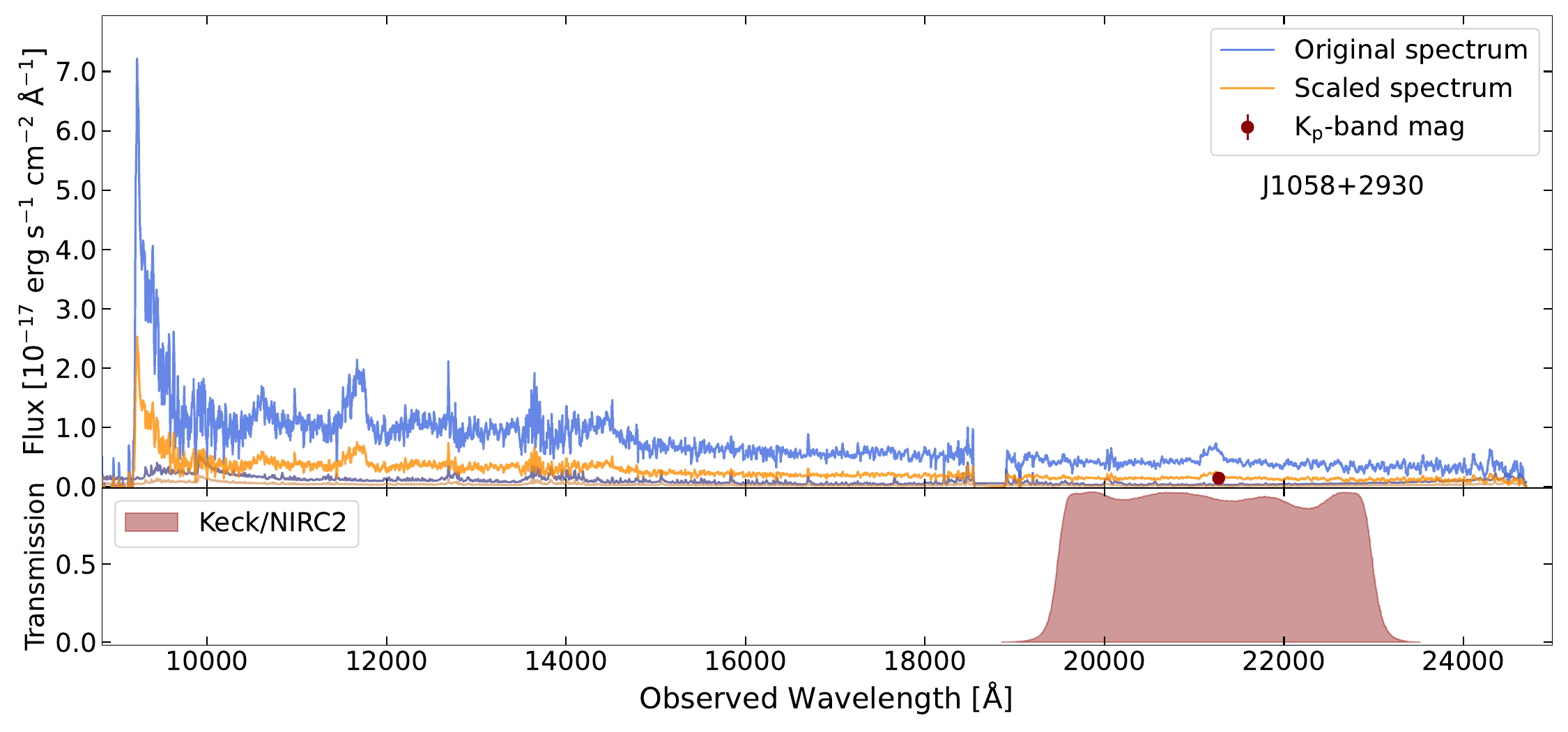}
    \caption{\textit{Top two panels}: Example cases of the spectra of J1342$+$0928 and J1007+2115 before (blue) and after (orange) applying the scale factor ($SF$) calculated from the J-band magnitudes of these quasars (as described in Equation \ref{eq:scalefac}), and the noise vectors. The lower plot shows (in brighter shades) the transmission curve of the passband filter used to get the photometry. We also include the Y, H, and K-band magnitudes, along with their respective transmission curves (in lighter shades), to compare with the synthetic photometry derived from the flux-scaled spectrum (magenta stars) and visually assess any discrepancies. \textit{Middle}: Same, but for J0923$+$0754 and its Y-band photometry. \textit{Bottom}: Same, but for J1058+2930 and its $\rm{K_{p}}$-band photometry.}
    \label{fig:scaled}
\end{figure*}

To correct the spectra, we scale them to a given magnitude in a specified passband filter.
First, we calculate the AB magnitude of the spectrum in a certain passband filter ($m_{\rm{spec}}$) by using the Python library \texttt{Speclite}\footnote{\url{https://speclite.readthedocs.io/en/latest/}}. Then we calculate a scale factor ($SF$) using the difference between the calculated magnitude from the spectrum and the true magnitude from the photometry ($m_{\rm{phot}}$), as shown in this formula that we get from the Pogson law:
\begin{equation}
\label{eq:scalefac}
    SF = 10^{0.4 (m_{\rm{spec}} - m_{\rm{phot}})}
\end{equation}
This scale factor is used to adjust the flux and inverse variance of the spectrum to match the true magnitude in the given passband (see Figure \ref{fig:scaled}).
For a subset of quasars, literature photometry is available not only in the J-band, but also in the Y, H, and K bands. We compute other synthetic magnitudes, but this time from the flux-scaled spectra and compare them to the additional photometry to assess any discrepancies. We find that the typical deviation between synthetic and observed magnitudes is around $10\%$, with a maximum discrepancy of approximately $40\%$. Two illustrative examples, one with high and one with low discrepancy, are shown in the top two panels of Figure \ref{fig:scaled}. We do not display synthetic photometry for the Y-band of the quasar J1342+0928 because the available spectrum does not fully cover the wavelength range of the Y-band filter ($9635-11025$ {\angstrom}). As a result, it is not possible to reliably compute the synthetic magnitude for this band.

\section{Absolute magnitude of J0910$-$0414 and J0923$+$0402 (BAL quasars)}\label{app:bal}

\begin{figure*}
    \centering
    \includegraphics[width=0.75\linewidth]{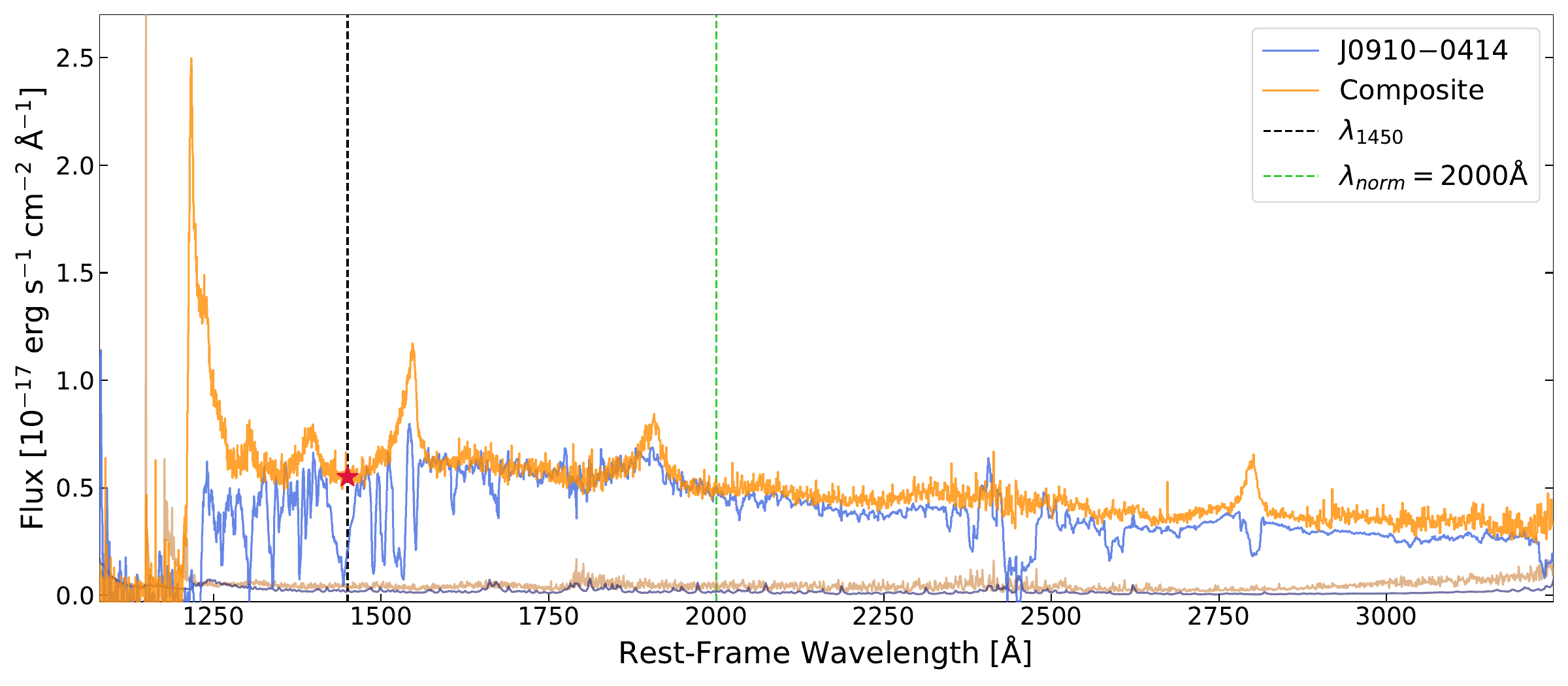}
    \includegraphics[width=0.75\linewidth]{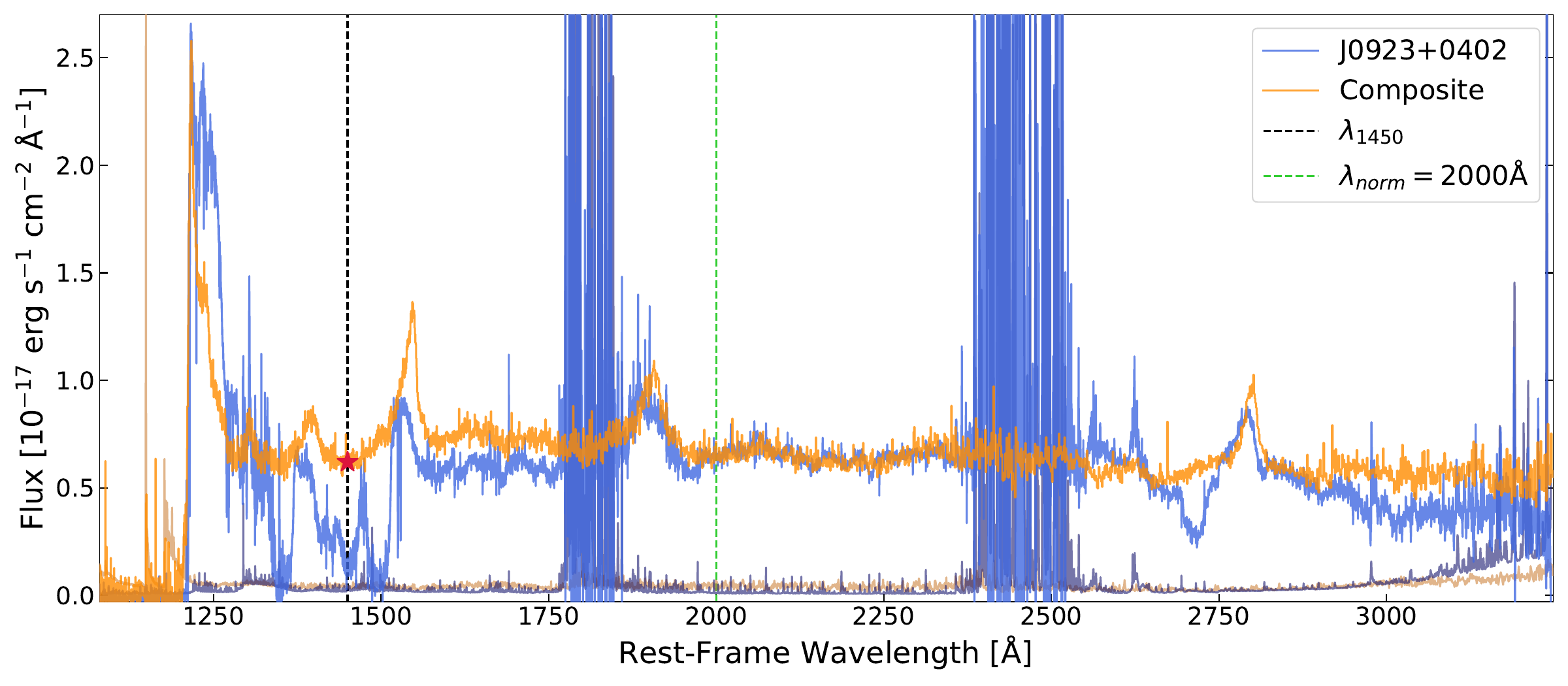}
    \caption{\textit{Top}: Match between the spectrum of J0910$-$0414 and the composite spectrum obtained from the sample excluding the BALs already corrected to get a better estimate of the $M_{1450}$. The spectrum of the BAL quasar is shown in blue, while the composite is in orange, with their noise vectors reported at the bottom of the plot. The back dashed line falls at $\lambda=1450$ {\angstrom} and the green dashed line is the normalization wavelength at $2000$ {\angstrom}. The magenta star is the "new" continuum level assumed for the BAL quasar at $1450$ {\angstrom}. \textit{Bottom}: Same, but for J0923$+$0402.}
    \label{fig:matchbal}
\end{figure*}

This Appendix describes the approach followed to calculate $M_{1450}$ in the two problematic cases of J0910$-$0414 and J0923$+$0402. These quasars are affected by strong BAL features, which mostly appear around the region at rest-frame $1450$ {\angstrom}, making the measurement of $M_{1450}$ with the method described in Section \ref{subsec:M1450} difficult to trust (note the black dashed line at $\lambda=1450$ {\angstrom} in Figure \ref{fig:matchbal} falling in the area affected by the BAL).
To correct this problem, we want to visualize what the continuum would be in a quasar with the same shape, but without BAL features. For this aim, we make a comparison using the composite spectrum created from the sample excluding the BAL quasars (described in Section \ref{sec:composite} and shown in Figure \ref{fig:composite}).
We start moving the observed spectrum (already flux-scaled using its J-band magnitude) to the rest-frame, while the composite spectrum is already in the rest-frame.
We scale the composite to the observed spectrum at a certain wavelength ($\lambda_{\rm{norm}}=2000$ {\angstrom}, in both cases, produces a good result) by multiplying for a factor $A$\footnote{It is the ratio between the flux of the BAL spectrum and the flux of the composite spectrum at the wavelength $\lambda_{\rm{norm}}$.}. We apply a power-law correction with an exponent $\alpha$, to rotate the composite and then have a better match with the orientation of the observed spectra, according to the following formula:
\begin{equation}
    f_{\rm{composite,new}} = f_{\rm{composite}} \cdot A \left( \frac{\lambda_{\rm{composite}}}{\lambda_{\rm{norm}}} \right)^\alpha
\end{equation}
We set $\alpha=0.3$ for J0910$-$0414 and $\alpha=0.8$ for J0923$+$0402.

Finally, we can calculate $M_{1450}$ (reported in Table \ref{tab:sample}) from the scaled composite spectrum following the method described in Section \ref{subsec:M1450}, assuming that the magenta star in Figure \ref{fig:matchbal} represents the "new" continuum level for the BAL quasar at $1450$ {\angstrom}. Figure \ref{fig:matchbal} shows the observed spectra in blue and the composite in orange, with their noise vectors reported at the bottom of each plot; the black dashed lines represent $\lambda=1450$ {\angstrom} and the green dashed lines are the normalization wavelengths, both set at $2000$ {\angstrom}.

\section{SNR}\label{app:snr}
\begin{figure}
    \includegraphics[width=\linewidth]{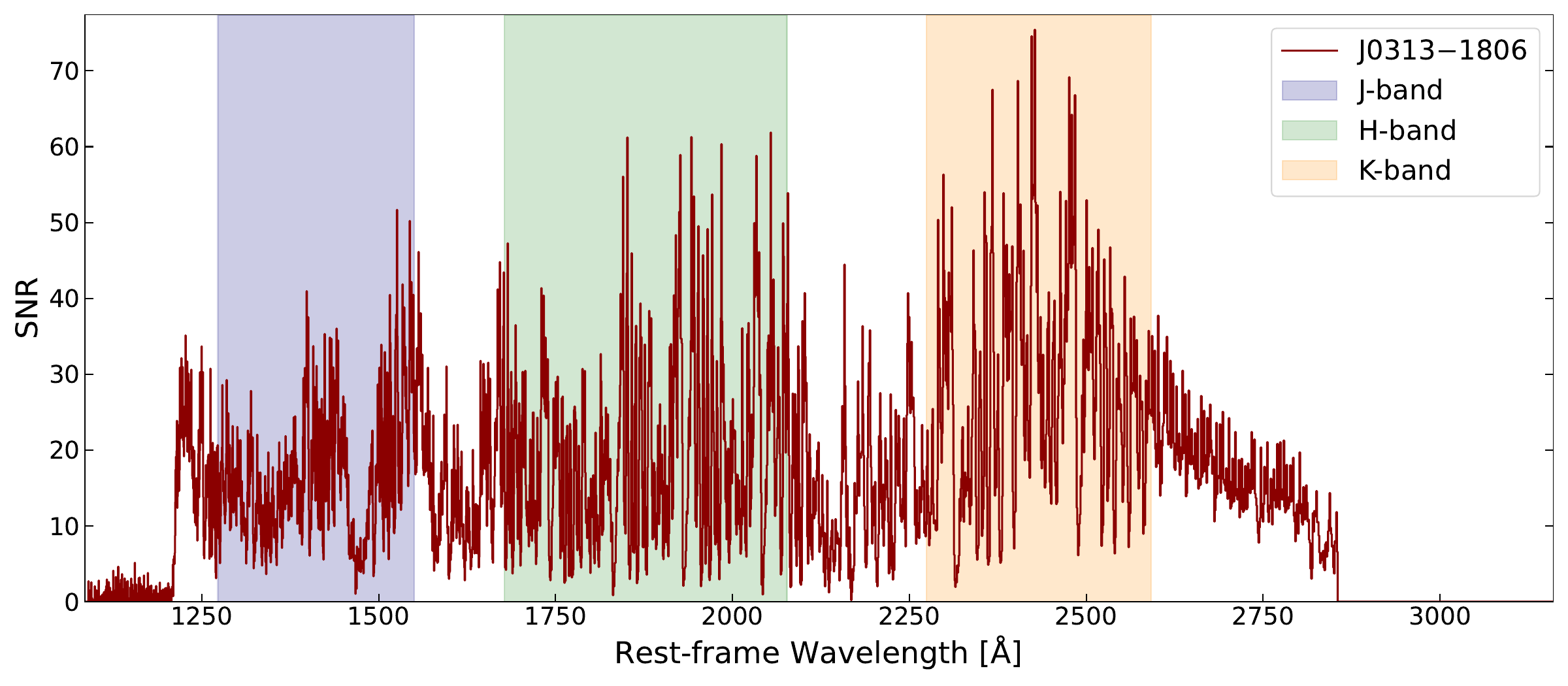}
    \includegraphics[width=\linewidth]{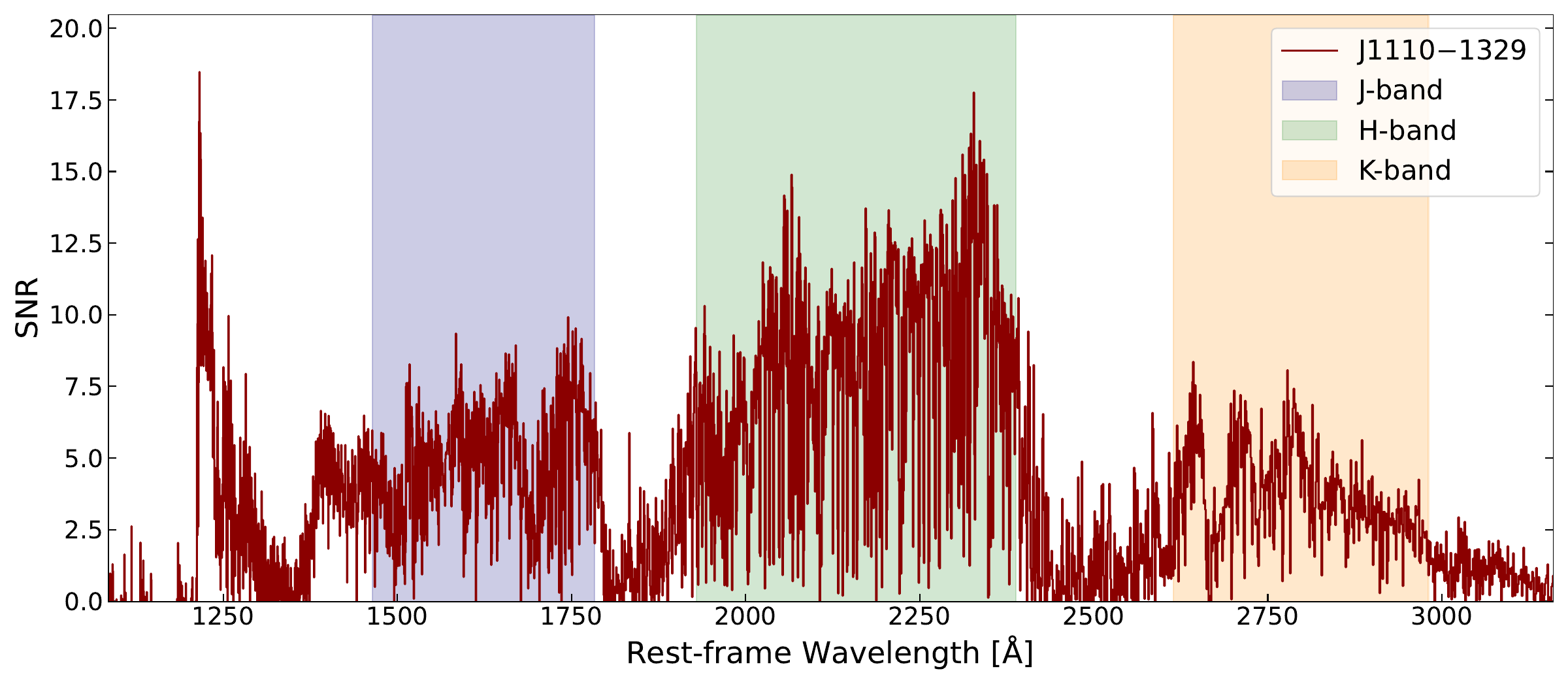}
    \caption{\textit{Top}: Example case of the distribution of the SNR along the wavelength grid [1040, 3332] {\angstrom} for the highest-$z$ quasar in the sample, J0313$-$1806 (dark red). The bands (J, H, and K) in which we compute the SNR are indicated according to different colors.\\
    \textit{Bottom}: Same, but for the lowest-$z$ quasar of the sample: J1110$-$1329.}
    \label{fig:snr}
\end{figure}

In Figure \ref{fig:snr}, we plot the distribution of the SNR along the rest-frame wavelength grid defined in Section \ref{sec:snr} for J0313$-$1806 and J1110$-$1329 (dark red curves), and the J, H, and K bands (indicated in Section \ref{sec:snr} too) are in different colors. These are two good example cases of spectra with medium/high and medium/low SNR, respectively.

\bsp	
\label{lastpage}
\end{document}